\documentclass{aa}  
\usepackage{graphicx}
\usepackage{txfonts}
\usepackage{longtable}
\usepackage{natbib}
\bibpunct{(}{)}{;}{a}{}{,}
\usepackage{url}
\begin{document} 
   \title{Very Long Baseline polarimetry and the gamma-ray connection in Markarian 421 during the broadband campaign in 2011}


   \author{R.~Lico\inst{1,2}\fnmsep\thanks{Email: rocco.lico@studio.unibo.it}, M.~Giroletti\inst{1}, M.~Orienti\inst{1,2}, J.~L.~G\'omez\inst{3}, C.~Casadio\inst{3}, F.~D'Ammando\inst{1,2}, M.~G.~Blasi\inst{1,4}, \\ W.~Cotton\inst{5}, P.~G.~Edwards\inst{6}, L.~Fuhrmann\inst{7}, S.~Jorstad\inst{8,9}, M.~Kino\inst{10}, Y.~Y.~Kovalev\inst{11,7},  T.~P.~Krichbaum\inst{7}, \\ A.~Marscher\inst{8}, D.~Paneque\inst{12}, B.~G.~Piner\inst{13} \and K.~V.~Sokolovsky\inst{11,14}.       
           }

   \institute{INAF Istituto di Radioastronomia, via Gobetti 101, 40129 Bologna, Italy
\and
Dipartimento di Fisica e Astronomia, Universit\`a di Bologna, via Ranzani 1, 40127 Bologna, Italy
\and
Instituto de Astrof\'{\i}sica de Andalucia, IAA-CSIC, Apdo. 3004, 18080 Granada, Spain
\and   
Scuola di Ingegneria, Universit\`a degli studi della Basilicata, viale dell'Ateneo Lucano 10, 85100 Potenza, Italy
\and   
National Radio Astronomy Observatory, Charlottesville, 520 Edgemont Road, VA 22903-2475, USA
\and
CSIRO Australia Telescope National Facility, PO Box 76, Epping NSW 1710, Australia
\and
Max-Planck-Institut f\"ur Radioastronomie, Auf dem H\"ugel 69, D-53121 Bonn, Germany
\and
Institute for Astrophysical Research, Boston University, 725 Commonwealth Avenue, Boston, MA 02215, USA
\and
Astronomical Institute, St. Petersburg State University, Universitetskij Pr. 28, 198504 St. Petersburg, Russia
\and
Korea Astronomy and Space Science Institute, 776 Daedeokdae-ro, Yuseong-gu, Daejeon 305-348, Republic of Korea
\and
Astro Space Center of Lebedev Physical Institute, Profsoyuznaya 84/32, 117997 Moscow, Russia
\and
Max-Planck-Institut f\"ur Physik, F\"ohringer Ring 6, D-80805 M\"unchen, Germany
\and
Department of Physics and Astronomy, Whittier College, 13406 E. Philadelphia Street, Whittier, CA 90608, USA
\and
Sternberg Astronomical Institute, Moscow State University, Universitetskij prosp. 13, 119992 Moscow, Russia
            }

   \date{Received June 05, 2014; accepted August 00, 2014}

 
  \abstract
   {This is the third paper in a series devoted to the analysis of the multiwavelength data from a campaign on the nearby ($z=0.03$) TeV blazar Mrk\,421 during 2011.}
   {We here investigate the structure of the high angular resolution polarization, the magnetic topology, the total intensity light curve, the $\gamma$-ray flux, and the photon index. We describe how they evolve and how they are connected.}
  {We analyzed data in polarized intensity obtained with the Very Long Baseline Array (VLBA) at twelve epochs (one observation per month from January to December 2011) at 15, 24, and 43\,GHz. For the absolute orientation of the electric vector position angles (EVPA) we used the D-terms method; we also confirm its accuracy. We also used $\gamma$-ray data from the {\it Fermi} Large Area Telescope on weekly time bins throughout 2011.
}
   {The source shows polarized emission, and its properties vary with time, frequency, and location along the jet. The core mean polarization fraction is generally between 1\% and 2\%, with a 4\% peak at 43\,GHz in March; the polarization angle is variable, mainly at 15\,GHz, where it changes frequently, and less so at 43\,GHz, where it oscillates in the range $114^\circ-173^\circ$. The jet polarization properties are more stable, with a fractional polarization of around 16\% and a polarization angle nearly perpendicular to the jet axis. 
   The average flux and photon index at $\gamma$-ray energies are (17.4 $\pm$ 0.5) $\times10^{-8}$ ph cm$^{-2}$ s$^{-1}$ and $\Gamma$ = 1.77 $\pm$ 0.02. 
The $\gamma$-ray light curve shows variability, with a main peak of $(38\pm11)\times10^{-8}$ ph cm$^{-2}$ s$^{-1}$ at the beginning of March and two later peaks centered on September 8 and November 13. The first $\gamma$-ray peak appears to be associated with the peak in the core polarized emission at 43\,GHz, as well as with the total intensity light curve. A discrete correlation function analysis yields a correlation coefficient of 0.54 at zero delay, with a significance level $> 99.7\%$.
}
   {
   With this multi-frequency study, we accurately determine the polarization properties of Mrk\,421, both in the core and in the jet region. The radio and $\gamma$-ray light curves are correlated. 
The observed EVPA variability at 15\,GHz is partly due to opacity and partly to a variable Faraday rotation effect. To explain the residual variability of the intrinsic polarization angle and the low degree of polarization in the core region, we invoke a blend of variable cross-polarized subcomponents with different polarization properties within the beam.}

\authorrunning{R. Lico et al.}
\titlerunning{VLBA Polarimetry and $\gamma$-ray connection in Mrk\,421 during 2011}

   \maketitle
%

\section{Introduction}

\begin{figure} 
 \centering
 \includegraphics[bb=0 45 595 667,width=0.73\columnwidth ,clip]{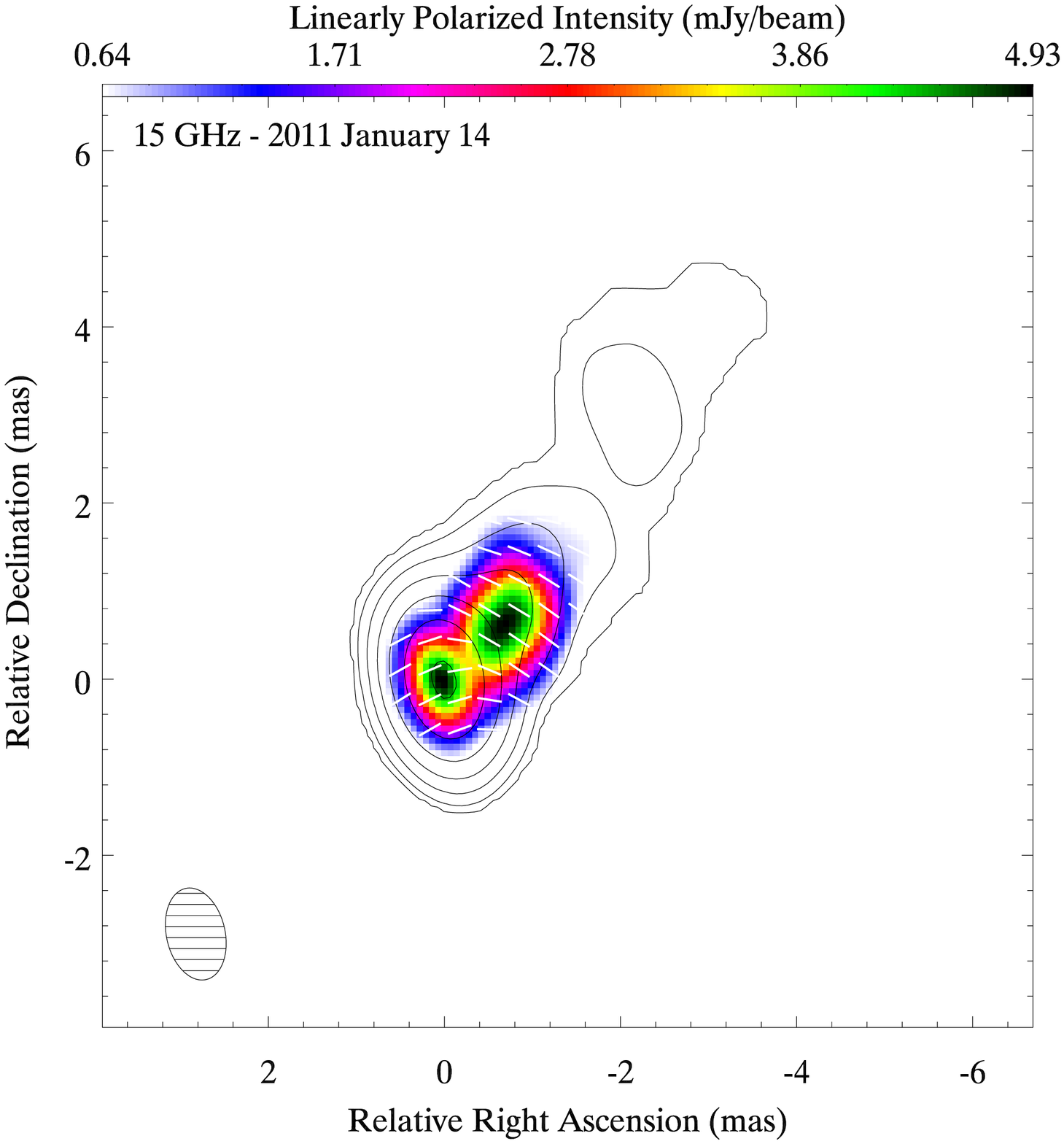} \\   
 \includegraphics[bb=0 45 595 693,width=0.73\columnwidth ,clip]{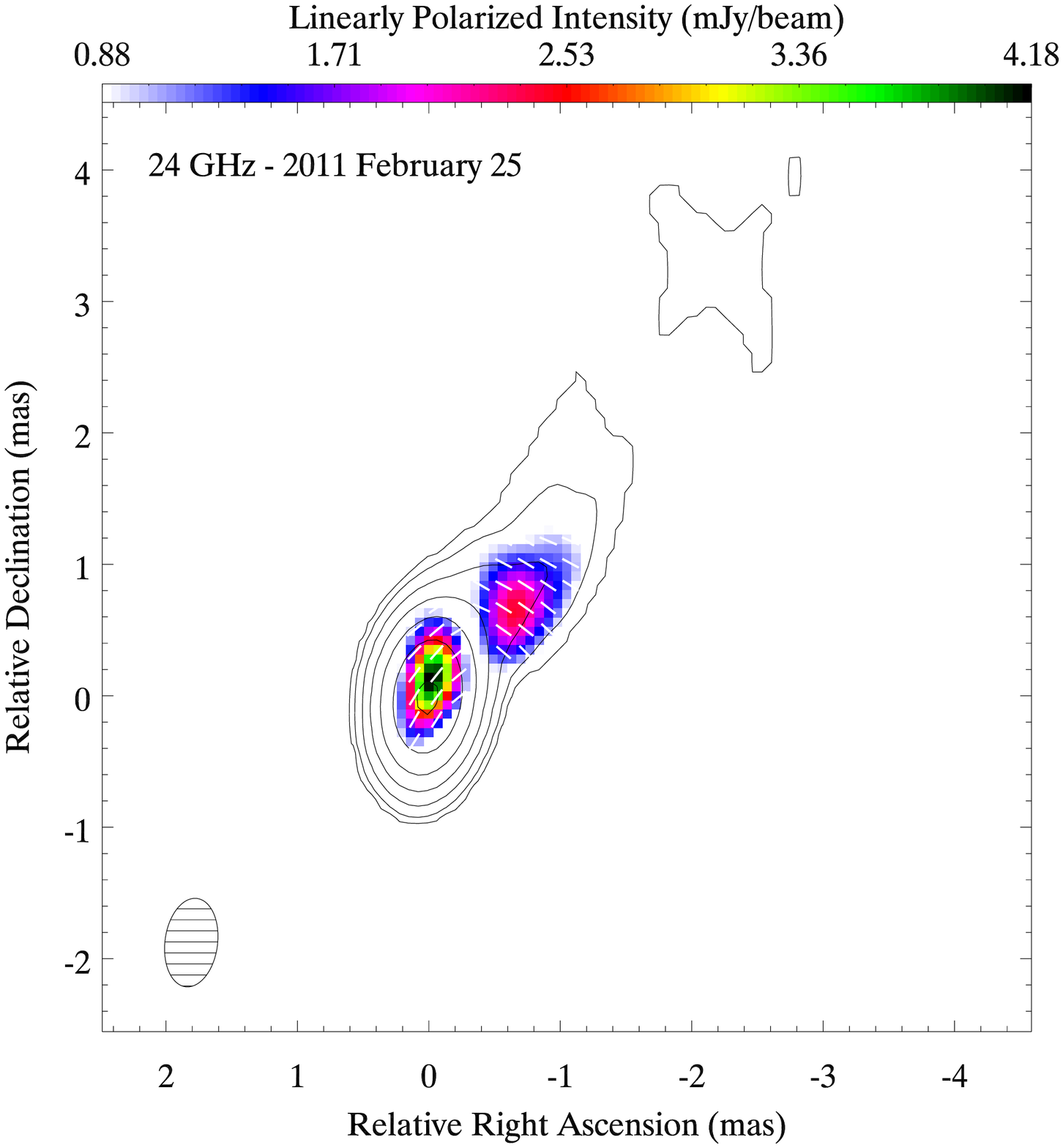} \\
 \includegraphics[bb=0 45 595 693,width=0.73\columnwidth ,clip]{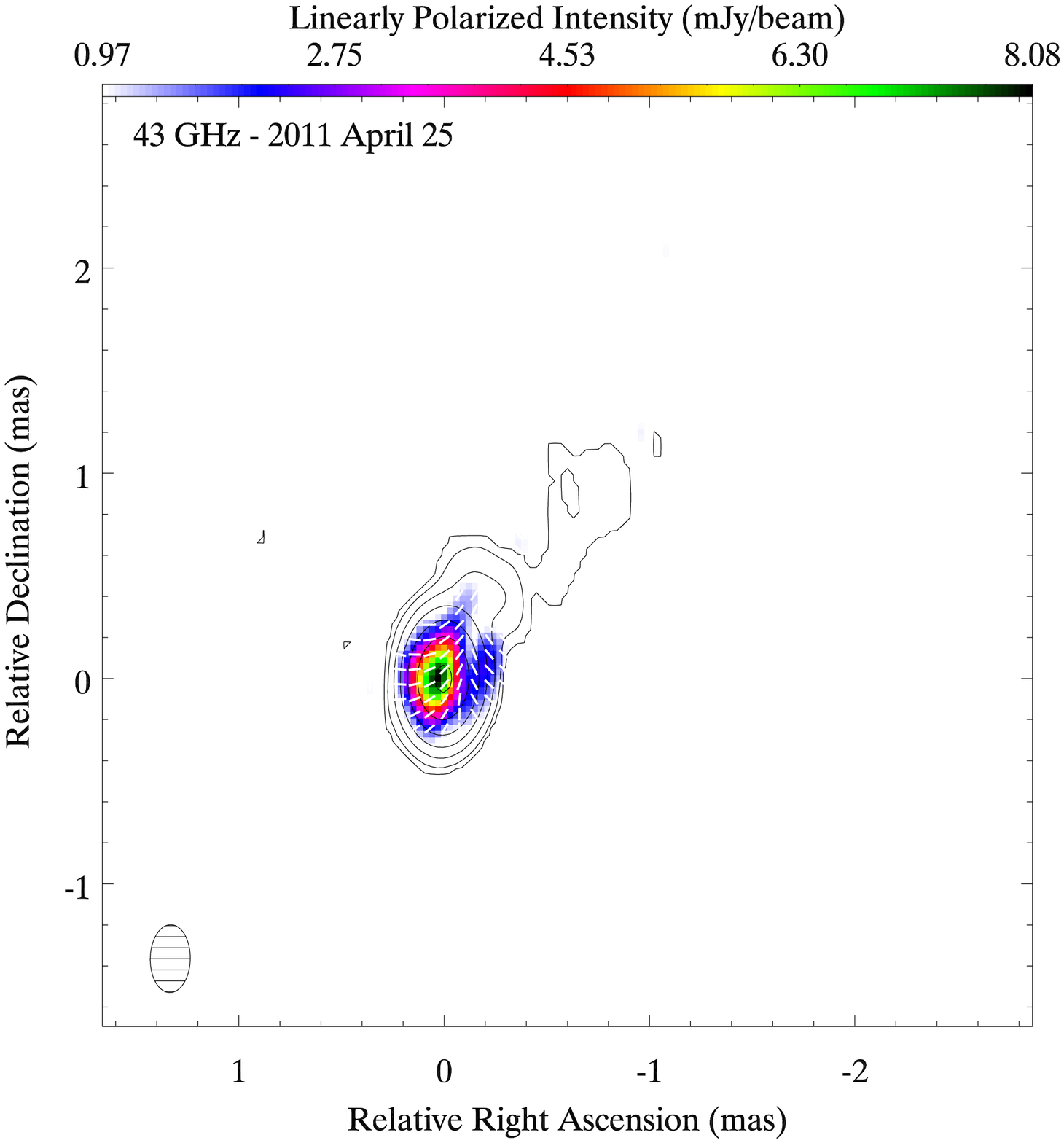} \\
\caption{Images of Mrk\,421 for the first observing epoch at 15\,GHz (top image), the second observing epoch at 24\,GHz (central image), and the third observing epoch at 43\,GHz (bottom image). Levels are drawn at $(-1, 1, 2, 4...) \times$ the lowest contour (that is, at 1.0 mJy/beam for 15 and 24\,GHz images and at 0.65 mJy/beam for the 43\,GHz image) increasing by factors of 2. The restoring beam, shown in the bottom left corner, has a value of 1.05~mas $\times$ 0.66~mas, 0.67~mas $\times$ 0.40~mas, and 0.33~mas $\times$ 0.19~mas for 15, 24, and 43\,GHz, respectively. The overlaid color maps show the linearly polarized intensity, and bars represent the absolute orientation of the EVPAs.}
\label{maps}
 \end{figure}

In the family of active galactic nuclei (AGN), blazars are the most powerful objects. Their relativistic jets are closely aligned with the line of sight. For this particular geometry, we observe that the radiation originates in the jet that points in our direction because of Doppler boosting effects. The emission in these objects is dominated by nonthermal radiation from relativistic electrons interacting with the magnetic field. 

Mrk\,421 is one of the nearest \citep[$z=0.03$,][]{deVaucouleurs1991} and brightest blazars and is therefore suitable for probing and investigating the physics of the innermost regions of relativistic jets. For this reason, Mrk\,421 has been intensively studied throughout the electromagnetic spectrum, especially since it was detected at TeV energies in 1992 \citep{Punch1992}. 

In general, the spectral energy distribution (SED) of blazars is dominated by the emission of the jet and consists of two separate components: a low-frequency hump, due to synchrotron emission by relativistic electrons within the jet, and a high-energy hump, that is commonly assumed to be inverse Compton (IC) scattering. In BL Lacs the high-energy IC component is commonly interpreted as synchrotron-self-Compton emission \citep[SSC, see][]{Abdo2011,Tavecchio2001}, resulting from the upscatter of the synchrotron photons off the jet electrons. For Mrk\,421, as for most TeV blazars \citep{Piner2013, Tiet2012}, the synchrotron hump peaks at soft X-rays, and for this reason, it is classified as a high-synchrotron-peaked (HSP) blazar \citep{Abdo2011}. 

Since the emission from these objects is dominated by nonthermal radiation, studying their polarization properties can provide important information on the magnetic field structure and the emission mechanisms. Furthermore, thanks to the multi-wavelength observations, we can investigate the location where the radiation is produced.

We observed Mrk\,421 with the Very Long Baseline Array (VLBA) at 15, 24, and 43\,GHz, both in total and polarized intensity. These datasets belong to a multi-frequency campaign, that was carried out during 2011, which also involved observations at submm (SMA), cm (e.g., F-GAMMA, Medicina), optical/IR (GASP), UV/X-ray ({\it Swift}, RXTE, MAXI), and $\gamma$-ray ({\it Fermi}, MAGIC, VERITAS) wavelengths.  

In two previous works we presented the complete analysis of observations in total intensity at 15 and 24\,GHz \citep{Lico2012} and at 43\,GHz \citep{Blasi2013}. We constrained some physical parameters such as the Doppler factor (in the radio emission region we found $\delta_r \sim 3$, while in the high-energy emission region we found $\delta_{\rm h.e.} \sim 14$), the viewing angle ($2^\circ < \theta < 5^\circ$), apparent speeds (no significant motion was detected in the jet), and the brightness temperature ($T_{\rm B,var} \sim 2.1\times 10^{10}$~K). 

In this work we present an analysis of the observations in linearly polarized intensity, which allows us to determine some physical parameters such as the degree of polarization and the absolute orientation of the electric vector position angle (EVPA), and to obtain some useful information on the magnetic field topology. We also used the data collected by the Large Area Telescope (LAT) onboard the {\em Fermi} satellite to produce $\gamma$-ray light curves for all of 2011. 

This paper is structured as follows: in Section 2 we describe the VLBA and {\em Fermi} observations and analysis, and we introduce the methods used for determining the EVPA orientation. In Section 3 we report the results of this work, and in Section 4 we discuss these results and interpret them in the astrophysical context. Throughout the paper we use the following conventions for cosmological parameters:  $H_0=70$ km sec$^{-1}$ Mpc$^{-1}$, $\Omega_M=0.25$ and $\Omega_\Lambda=0.75$, in a flat Universe. 
All angles are measured from north through east. At the redshift of the target, 1 mas corresponds to 0.59 pc.

\begin{table*}
\caption{Final EVPA rotations for 15, 24, and 43\,GHz (Cols. 3, 6, and 9, respectively); numbers in boldface refer to the comparison with JVLA values. We also report the relative rotations, obtained by comparing antenna tables for consecutive epochs, and the reference antennas used for the phase calibration.}
\label{tabrotations}      
\centering
\tiny 
\begin{tabular}{ccccccccccc}
\hline
\hline
Epoch & MJD & Final  $\Delta$\tablefootmark{a} & $\Delta$D-terms\tablefootmark{b} & Reference & Final  $\Delta$\tablefootmark{a} & $\Delta$D-terms\tablefootmark{b} & Reference & Final  $\Delta$\tablefootmark{a} & $\Delta$D-terms\tablefootmark{b} & Reference\tablefootmark{c} \\
year/month/day & & (deg) & (deg) & antenna & (deg) & (deg) & antenna & (deg) & (deg) & antenna \\
\hline
 & & $15\,GHz$ & & & $24\,GHz$ & & & $43\,GHz$ & & \\
\hline
2011/01/14 & 55575 & $-$21.7         &    &  PT & $-$8.5          &         & PT & $-$3.2  &   & PT	 \\
2011/02/25 & 55617 & $-$21.7         &  0 &  PT & $-$8.5          & 0       & PT & $-$3.2  & 0 & PT \\
2011/03/29 & 55649 & \textbf{$-$21.7}&  0 &  PT & \textbf{$-$8.5} & 0       & PT & \textbf{$-$3.2}  & 0 & PT \\
2011/04/25 & 55675 & $-$21.7         &  0 &  PT & $-$8.5          & 0       & PT & $-$3.2  & 0 & PT \\
2011/05/31 & 55712 & $-$21.7         &  0 &  PT & $-$8.5          & 0       & PT & $-$3.2  & 0 & PT \\
2011/06/29 & 55741 &  \textbf{22.2}  & 45 &  OV & $-$91           & $-$82.5 & OV & $-$25.2 & $-$22 & KP\\
2011/07/28 & 55770 &  22.2           &  0 &  OV & $-$153.5        & $-$62.5 & KP & $-$25.2 & 0 & KP\\
2011/08/29 & 55802 &  85.2           & 63 &  KP & $-$153.5        & 0       & KP & $-$25.2 & 0 & KP\\
2011/09/28 & 55832 & 157.2           & 72 &  PT & $-$6.5          & $-$33   & PT & 6.8   & 32 & PT\\
2011/10/29 & 55863 & 157.2           &  0 &  PT & $-$6.5          & 0       & PT & 6.8   & 0 & PT\\
2011/11/28 & 55893 &  \textbf{25.5}  & 45 &  OV & \textbf{$-$79.7}& $-$67.7 & OV & \textbf{$-$49.2} & $-$51 & OV\\
2011/12/23 & 55918 &  $-$19.5        &-45 &  PT & $-$9.7          & $-$110  & PT & 0.8   & 50 & KP\\
\hline
\hline\\
\end{tabular}
\tablefoot{
\begin{tiny}
\newline
\tablefoottext{a}{Final rotation to apply to obtain the absolute EVPA orientation.}\\
\tablefoottext{b}{Relative rotation obtained by comparing antenna table of two consecutive epochs.}\\
\tablefoottext{c}{PT = Pie Town, KP = Kitt Peak, OV = Owens Valley.}\\
\end{tiny}
}
\end{table*}

\begin{figure}
\includegraphics[bb=77 360 539 720, width=1.0\columnwidth ,clip]{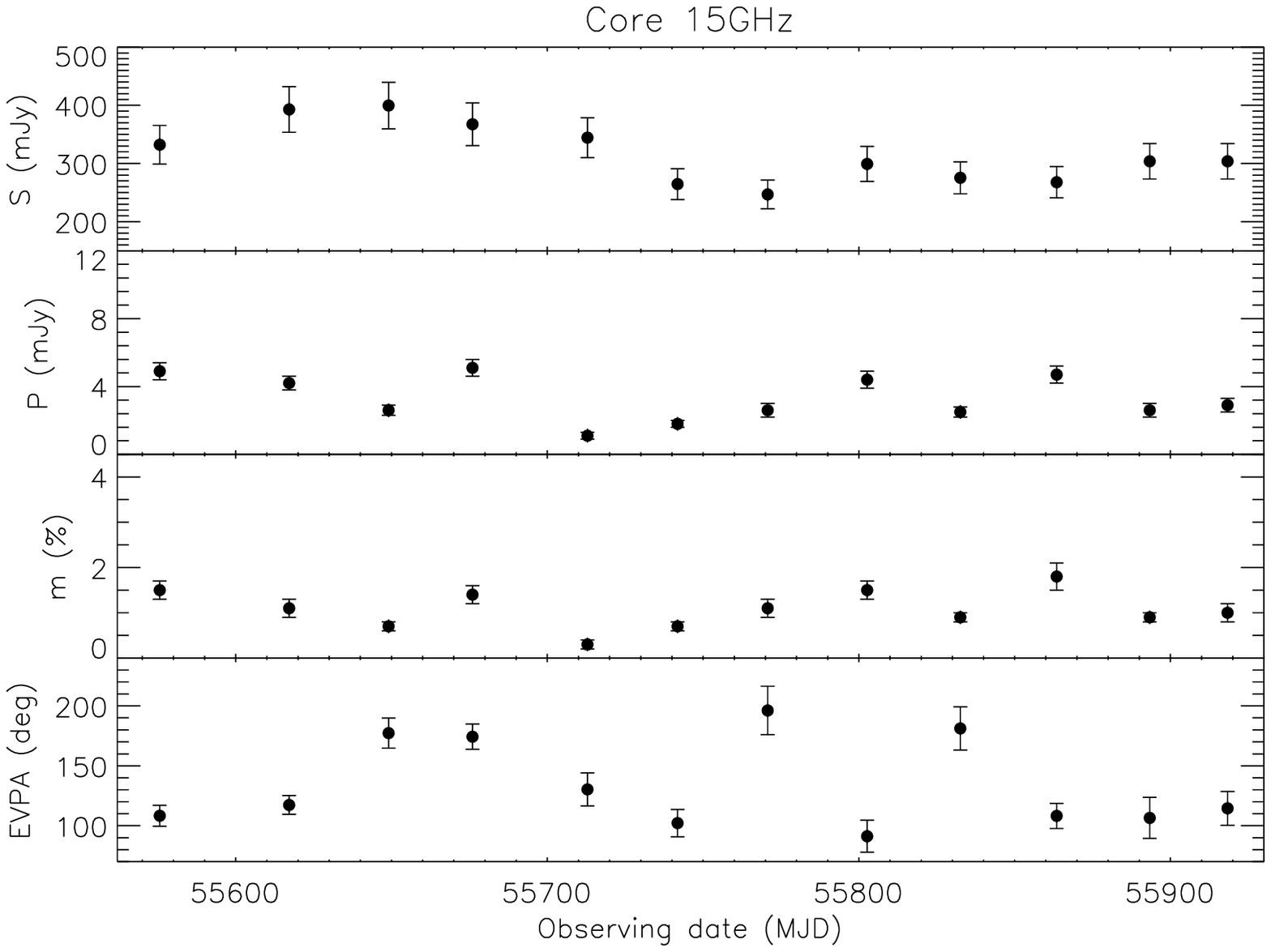} \\    
\includegraphics[bb=77 360 539 720, width=1.0\columnwidth ,clip]{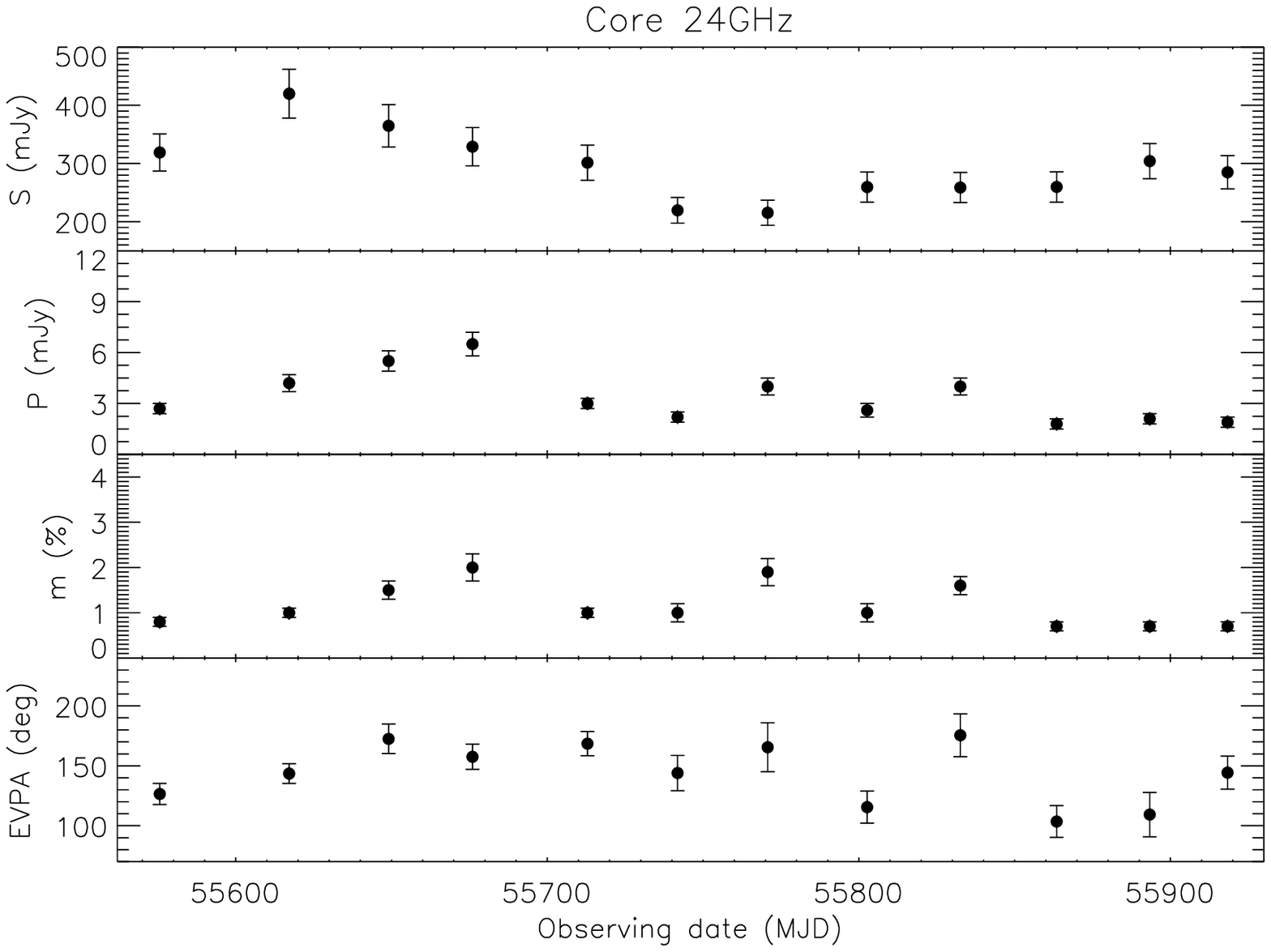} \\
\includegraphics[bb=77 360 539 720, width=1.0\columnwidth ,clip]{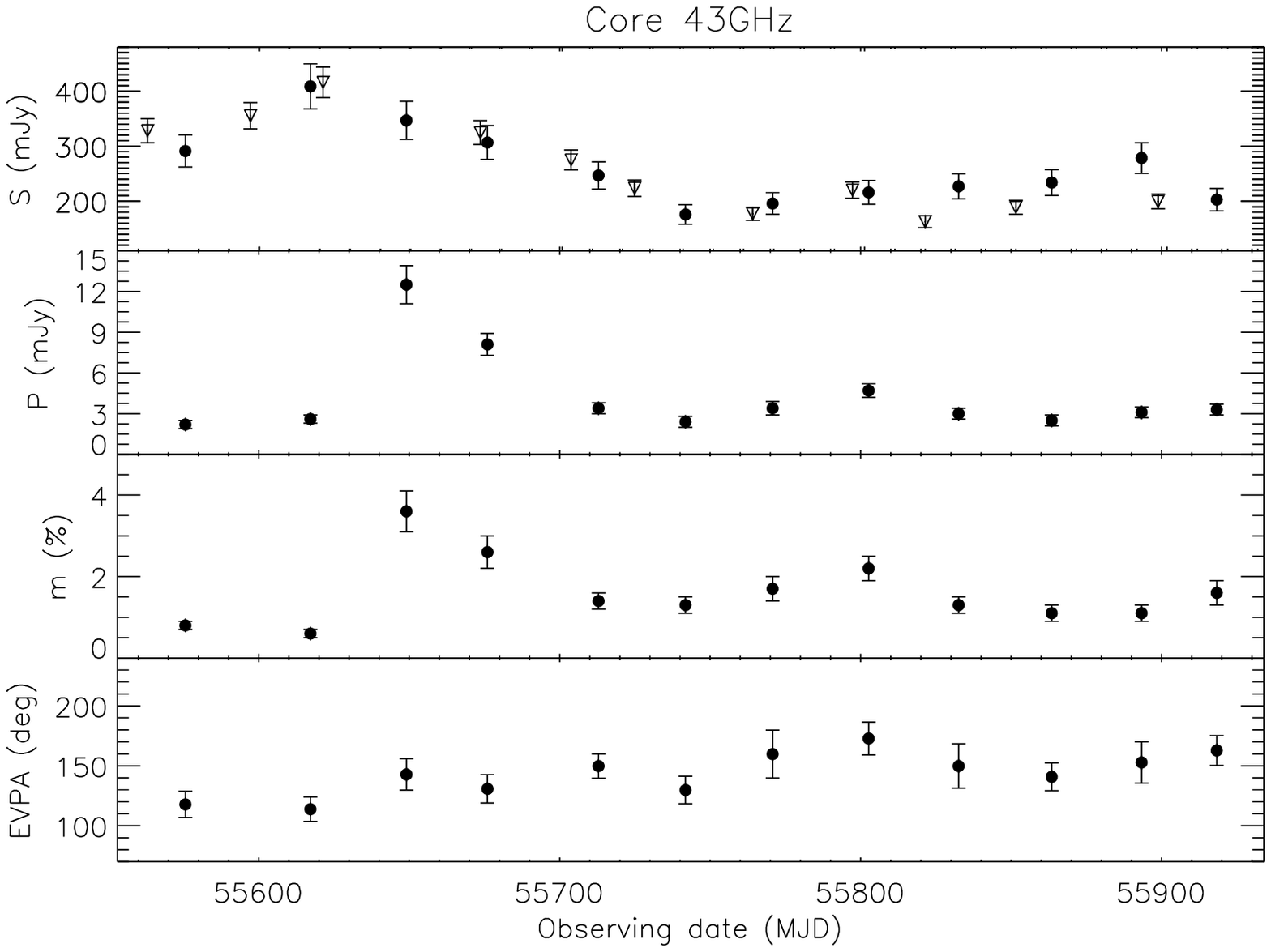} \\
\caption{Evolution with time of some physical parameters of Mrk\,421 in the core region at 15\,GHz (upper frame), 24\,GHz (middle frame), and 43\,GHz (lower frame). For each frame we report from the first to the fourth panel the light curves for the total intensity, the polarized flux density, the fractional polarization, and the EVPA values. Triangles in the lower frame (first panel) represent the VLBA 43\,GHz observations of Mrk\,421 provided by the VLBA-BU-BLAZAR program.} 
\label{plots_core}
\end{figure}

\section{Observations and data reduction}

\subsection{VLBA data: details and analysis}
We observed Mrk\,421 with the VLBA once per month throughout 2011 at three frequencies: 15, 24, and 43\,GHz. Observations were carried out in total and polarized intensity (in right and left circular polarization). For the details of the observations at 15\,GHz and 24\,GHz, see \citet{Lico2012}, and for the 43\,GHz observations, see \citet{Blasi2013}. Our 43\,GHz dataset has been expanded by adding data from 11 epochs provided by the VLBA-BU-BLAZAR program\footnote{\url{http://www.bu.edu/blazars/VLBAproject.html}} (see Table~\ref{bostondata}). We updated the total intensity flux densities here with respect to our previous works: for the main campaign datasets, we made a new and independent calibration (for the sake of a homogeneous polarization calibration, see below), which resulted in new values entirely consistent with the previous ones within the uncertainties; for the supplementary Boston University datasets, the flux density scale was adjusted by a scaling factor obtained from comparing of the flux densities between the VLBA and the single-dish Mets\"ahovi near-simultaneous datasets, resulting in a significantly improved accuracy of the flux density scale.

For the calibration, the fringe-fitting procedure, and the detection of cross-polarized fringes we used the software package Astronomical Image Processing System (AIPS) \citep{Greisen2003}. We produced the cleaned and final images with the software package DIFMAP \citep{Shepherd1997}. The polarization D-terms were determined with the task LPCAL in AIPS. We compared antenna tables (which provide the phase and amplitude values for the R and L circular polarization for each antenna) at two consecutive epochs to determine the relative EVPA rotations using the IDL routines developed by J.~L.~G\'omez.

We calibrated the instrumental polarization using the strong (flux density $> 1$ Jy) and structureless source J1310+3220.  This source also provides good coverage of the parallactic angle ($>100^\circ$) and has negligible polarization on large scales, which is important because of the very different angular resolution of the VLBA with respect to the Karl G. Jansky Very Large Array (JVLA).

\subsection{Polarization calibration}
\label{methods}
In general, to obtain the absolute orientation of the EVPAs (defined as $\chi =0.5 \times \arctan(U/Q)$, where $U$ and $Q$ are Stokes flux densities) in very long baseline interferometric (VLBI) observations, a comparison with quasi-simultaneous single-dish or JVLA observation is required. This is because of the lack of polarization calibrators with stable EVPAs on (sub)milliarcsecond scales. 

To determine the EVPA absolute orientation, we used the method developed by \citet{Leppanen1995}, which makes use of the instrumental polarization parameters (the so-called D-terms). This method provides us with an independent way of calibrating the absolute right-left (R-L) circular polarization phase offset; it is based on the assumption that the D-terms change slowly with time \citep[see][]{Gomez2002}. For most of the antennas we found that D-terms remain stable during the whole 12-month observing period.

The method consists of comparing the D-terms for each antenna in consecutive epochs, which yields the relative rotation in right (R) and left (L) circular polarization. In other words, the phase offset in R and L between two epochs is provided by the phase difference of the D-terms. 
We note that since the VLBA consists of ten antennas, twenty different values are involved in the comparison to determine the relative rotation between two epochs,  (two R-L values for each antenna). This guarantees an accurate and reliable determination of the phase offset in case of an antenna failure. 

The relative rotations obtained for 15, 24, and 43\,GHz are reported in Cols. 4, 7, and 10 in Table~\ref{tabrotations}. For a specific epoch the relative rotation may differ at different frequencies because it depends on the reference antenna used for the calibration; the reference antennas used at the different frequencies are listed in Cols. 5, 8, and 11 in the same table. 

After determining the relative rotations, we set the absolute EVPA calibration for one epoch by comparison with a JVLA observation provided by the POLCAL program\footnote{\url{http://www.aoc.nrao.edu/~smyers/evlapolcal/polcal_master.html}} (values in boldface in Table~\ref{tabrotations}). Then we determined the absolute orientation for all the EVPAs by applying the relative rotations obtained from the D-terms. For example, after fixing a value in the third column for the 15\,GHz data by the comparison with JVLA, we obtained the other values by summing the previous value in the same column and the relative rotation for the same epoch in Col. 4. For example, at 15\,GHz in Col. 3 in Table~\ref{tabrotations} for the 11th observing epoch (November 2011) we obtain the value of $25.5^\circ$ by the comparison with JVLA. Then, to obtain the absolute rotation for the consecutive epoch (December 2011), we just add the relative rotation to the value of $25.5^\circ$ (fourth column), which in this case is $-45^\circ$, and we obtain a final rotation of $-19.5^\circ$. The relative rotation between two epochs is $0^\circ$ for the same reference antenna at both epochs. 

At 15\,GHz we have three JVLA measurements taken during 2011, which enabled two cross checks. The values obtained with the D-terms method and those from the comparison with the JVLA agree within $5^\circ$. For the 24 and 43\,GHz observations we have two JVLA measurements, which we also cross-checked, finding very good agreement within $5^\circ$. This confirms the validity and accuracy of the D-terms calibration method reported previously by \citet{Gomez2002}.

\subsection{Determination of uncertainties}
Error bars for the total intensity flux density ($S$) and the linearly polarized emission (defined as $P=\sqrt{Q^2+U^2}$) were calculated by considering a calibration uncertainty $\sigma_c$ of about $10\%$ of the flux density and a statistical error provided by the map rms noise. 

To determine the statistical error for the jet emission we also took the number of beams into account:

\begin{equation}
\label{err_flux}
\Delta_S= \sqrt{\sigma_c^2 + \left(\sqrt{\frac{\mbox{box size}}{\mbox{beam size}}} \times rms\right)^2}.
\end{equation} 
The box size term is defined in Sect.~\ref{morphology}. 
The uncertainties in fractional polarization (defined as $m=P/S$) were calculated from error propagation theory:
\begin{equation}
\Delta m = \frac{1}{S} \sqrt{\sigma_P^2 + \left(\frac{P}{S} \times \sigma_S\right)^2},
\end{equation}
where $\sigma_S$ and $\sigma_P$ represent the uncertainties in $S$ and $P$.

The uncertainties in EVPA values were calculated by taking into account all of these contributions:
\begin{equation}
\Delta \chi = \sqrt{\sigma_\mathrm{cal}^2+\sigma_\mathrm{D-terms}^2+\sigma_\mathrm{JVLA}^2+\sigma_{\chi}^2},
\end{equation}
where $\sigma_\mathrm{cal}$ is the scatter of the value measured on the polarization map, $\sigma_\mathrm{D-terms}$ is the calibration uncertainty introduced when we compare D-terms at different epochs to obtain the relative rotations by using the antenna tables, and $\sigma_\mathrm{JVLA}$ is the $5^\circ$ mean difference between the JVLA and the D-terms methods. The last term $\sigma_{\chi}$ is the uncertainty in $\chi$ calculated from error propagation theory:
\begin{equation}
\sigma_{\chi} = \frac{0.5}{Q^2+U^2} \sqrt{U^2 \Delta_{Q}^2 + Q^2 \Delta_{U}^2},
\end{equation}
where $\Delta_Q$ and $\Delta_U$ are the uncertainties in the $Q$ and $U$ Stokes flux densities, which are calculated using Eq.~(\ref{err_flux}). 
Since $\Delta_Q \sim \Delta_U$ \citep{fanti2001}, the formula becomes
\begin{equation}
\sigma_{\chi} = \frac{0.5 \times \Delta_Q }{\sqrt{Q^2+U^2}} = 0.5 \times \frac{\Delta_Q }{P}.
\end{equation}

To determine the polarization parameters we did not include the random noise correction \citep{wardle1974}; this contribution is always within the uncertainties.

\subsection{{\em Fermi}-LAT data: selection and analysis}
The {\em Fermi}-LAT  is a pair-conversion telescope operating from 20 MeV to $>$ 300 GeV. Further details about the {\em Fermi}-LAT are given in \citet{Atwood2009}. The LAT data reported here were collected from 2011 January 1 (MJD 55562) to December 31 (MJD 55926). During this time, the {\em Fermi} observatory operated almost entirely in survey mode. The analysis was performed with the software package \texttt{ScienceTools} version v9r32p5. The LAT data were extracted within a region of $20^{\circ}$ radius centered on the location of Mrk\,421. Only events belonging to the `Source' class were used. The time intervals collected when the rocking angle of the LAT was greater than 52$^{\circ}$ were rejected. In addition, a cut on the zenith angle ($<100^{\circ}$) was applied to reduce contamination from the Earth limb $\gamma$ rays, which are produced by cosmic rays interacting with the upper atmosphere. 
The spectral analysis was performed with the instrument response functions  \texttt{P7REP\_SOURCE\_V15} using an unbinned maximum-likelihood method implemented  in the Science tool \texttt{gtlike}. A Galactic diffuse emission model and isotropic component, which is the sum of an extragalactic and residual cosmic-ray background, were used to model the background\footnote{\url{http://fermi.gsfc.nasa.gov/ssc/data/access/lat/BackgroundModels.html}}. 
The normalizations of the two components in the background model were allowed to vary freely during the spectral fitting. 

\begin{table}
\caption{VLBA 43\,GHz data provided by the Boston University blazar monitoring program.}
\label{bostondata}     
\centering    
\tiny                 
\begin{tabular}{c c c c}        
\hline\hline
Epoch & MJD & S\tablefootmark{a} & $\sigma_S$\tablefootmark{b} \\
year/month/day & & (mas) & (mas) \\
\hline                        
2011/01/02 & 55563 & 328.0 & 21.9 \\
2011/02/04 & 55597 & 355.3 & 23.7 \\
2011/03/01 & 55621 & 415.9 & 27.7 \\
2011/04/21 & 55673 & 324.6 & 21.6 \\
2011/05/22 & 55703 & 275.0 & 18.3 \\
2011/06/12 & 55724 & 223.6 & 14.9 \\
2011/07/21 & 55763 & 177.0 & 11.8 \\
2011/08/23 & 55796 & 220.3 & 14.7 \\
2011/09/16 & 55820 & 162.7 & 10.8 \\
2011/10/16 & 55850 & 189.0 & 12.6 \\
2011/12/02 & 55897 & 199.6 & 13.3 \\
\hline 
\hline
\end{tabular}
\tablefoot{
\begin{tiny}
\newline
\tablefoottext{a}{Flux density in mJy.}\\
\tablefoottext{b}{Estimated errors for the flux density.}\\
\end{tiny}
}
\end{table}

We analyzed a region of interest of $10^{\circ}$ radius centered on the location of Mrk\,421.
We evaluated the significance of the $\gamma$-ray signal from the sources by means of the maximum-likelihood test statistic TS = 2$\Delta$log(likelihood) between models with and without a point source at the position of Mrk\,421 \citep{Mattox1996}. The source model used in \texttt{gtlike} includes all of the point sources from the second {\em Fermi}-LAT catalog \citep[2FGL;][]{Nolan2012} as well as in a preliminary third {\em Fermi}-LAT catalog (Ackermann et al., in prep.) that fall within $15^{\circ}$ of the source. 
The spectra of these sources were parametrized by power-law functions, except for  2FGL\,J1015.1+4925, for which we used a log-parabola as in the 2FGL catalog. A first maximum-likelihood analysis was performed to remove from the model the sources with TS $<$ 25 and/or the predicted number of counts based on the fitted model $N_{pred} < 3$. A second maximum-likelihood analysis was performed on the updated source model. In the fitting procedure, the normalization factors and the photon indices of the sources lying within 10$^{\circ}$ of Mrk\,421 were left as free parameters. For the sources located between 10$^{\circ}$ and 15$^{\circ}$ from our target, we kept the normalization and the photon index fixed to the values from the 2FGL catalog. 

The systematic uncertainty on the effective area\footnote{\url{http://fermi.gsfc.nasa.gov/ssc/data/analysis/LAT_caveats.html}} is $10\%$ below 100 MeV \citep{Ackermann2012}, decreasing linearly with the logarithm of energy to $5\%$ between 316 MeV and 10 GeV, and increasing linearly with the logarithm of energy up to $15\%$ at 1 TeV.

\begin{figure}
\includegraphics[bb=77 360 539 720, width=1.0\columnwidth ,clip]{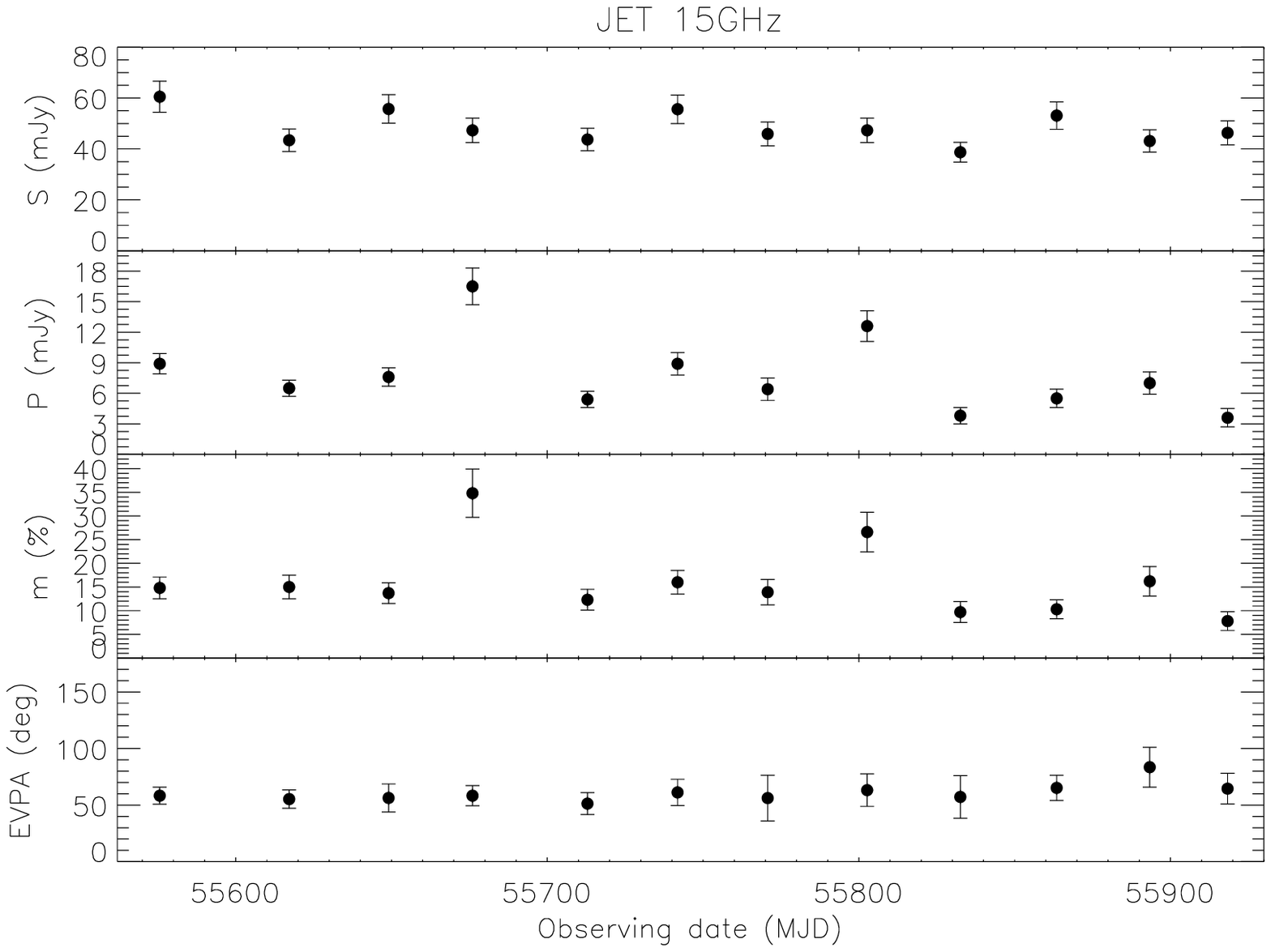} \\
\includegraphics[bb=77 360 539 720, width=1.0\columnwidth ,clip]{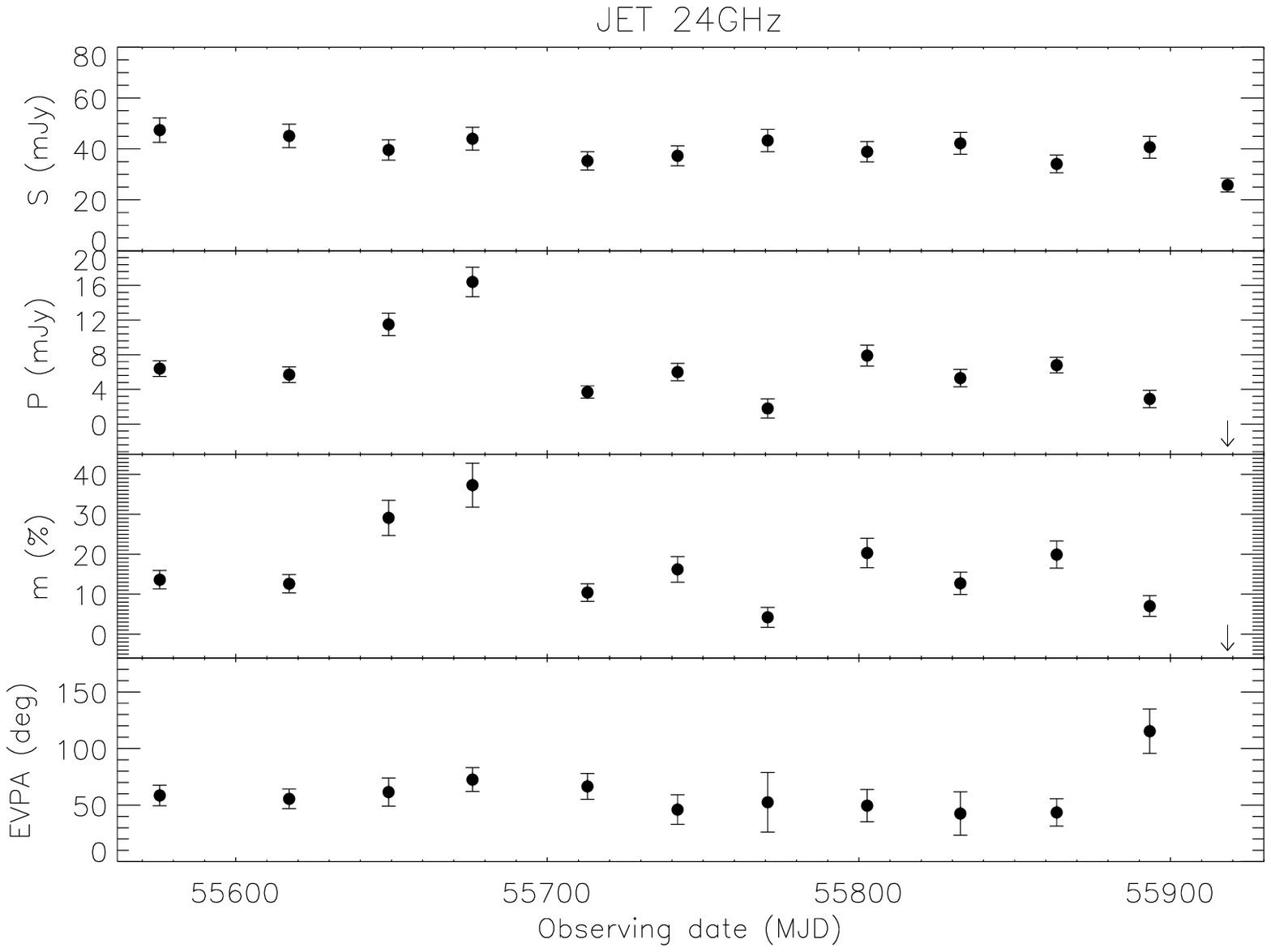} \\
\caption{Evolution with time of some physical parameters in the jet region at 15\,GHz (upper frame) and at 24\,GHz (lower frame). For each frame we report from the first to the fourth panel the light curves for the total intensity, the polarized flux density, the fractional polarization, and the EVPA values.} 
\label{plots_jet}
\end{figure}

\begin{figure*}
\includegraphics[width=0.67\columnwidth]{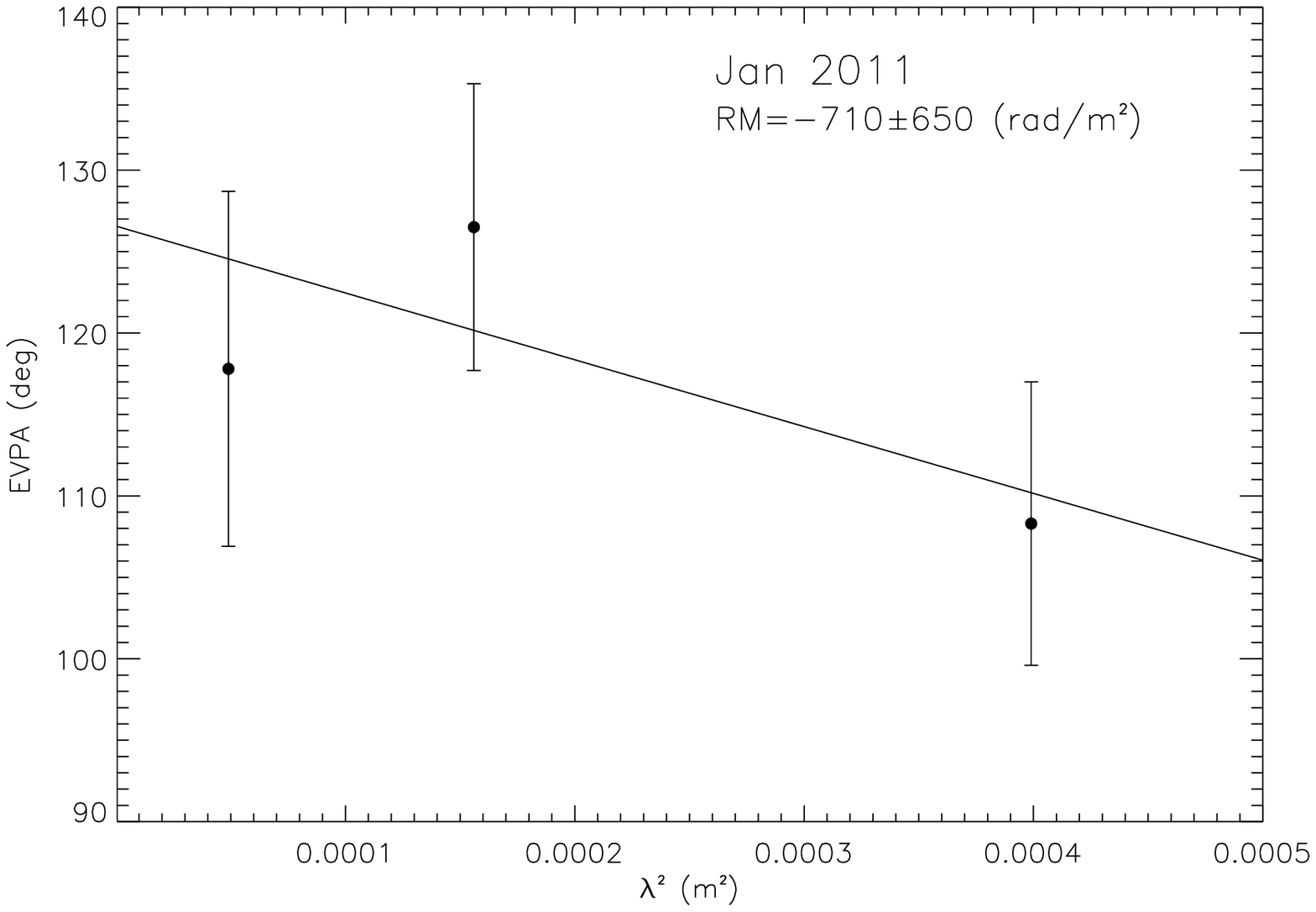}
\includegraphics[width=0.67\columnwidth]{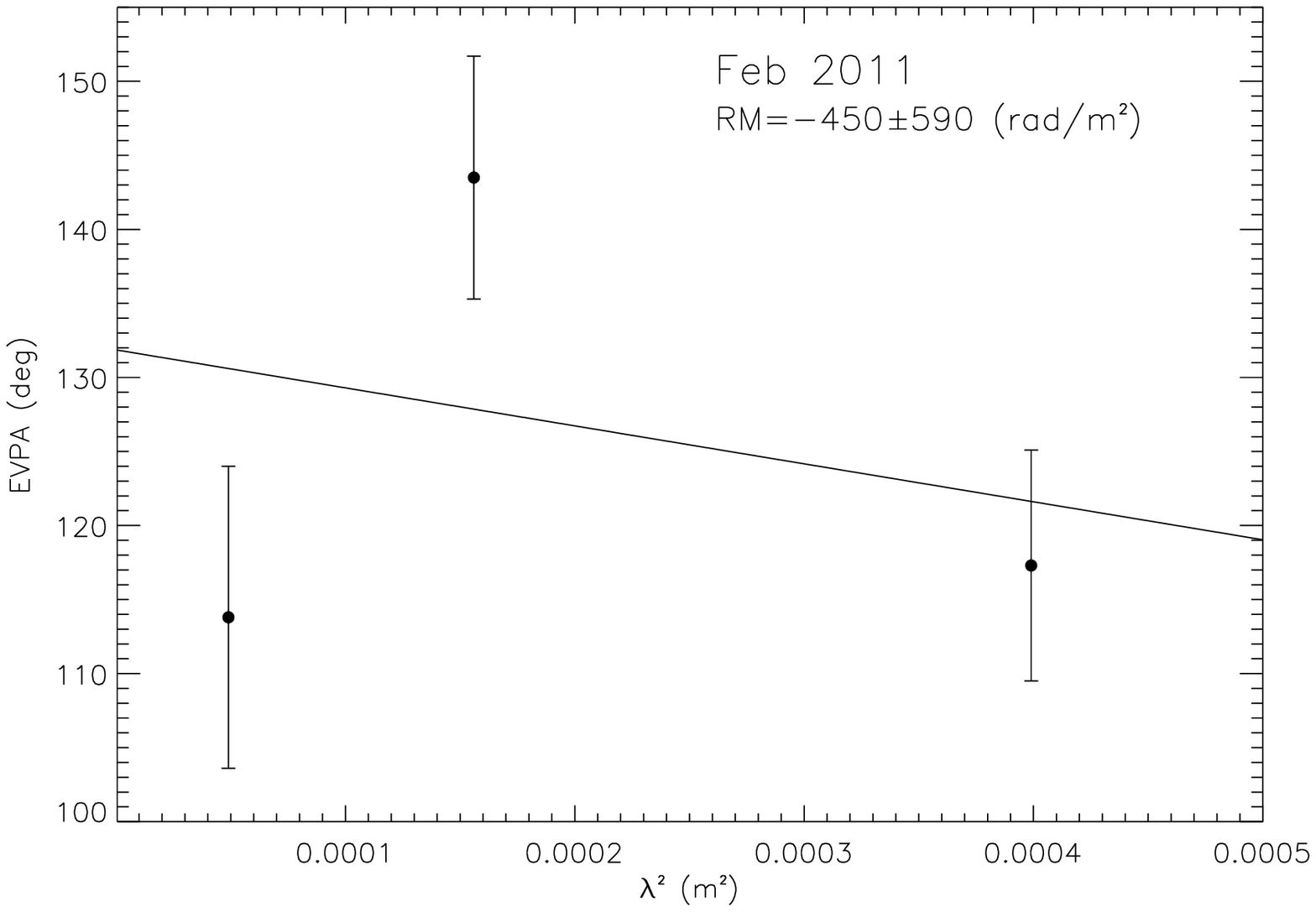}
\includegraphics[width=0.67\columnwidth]{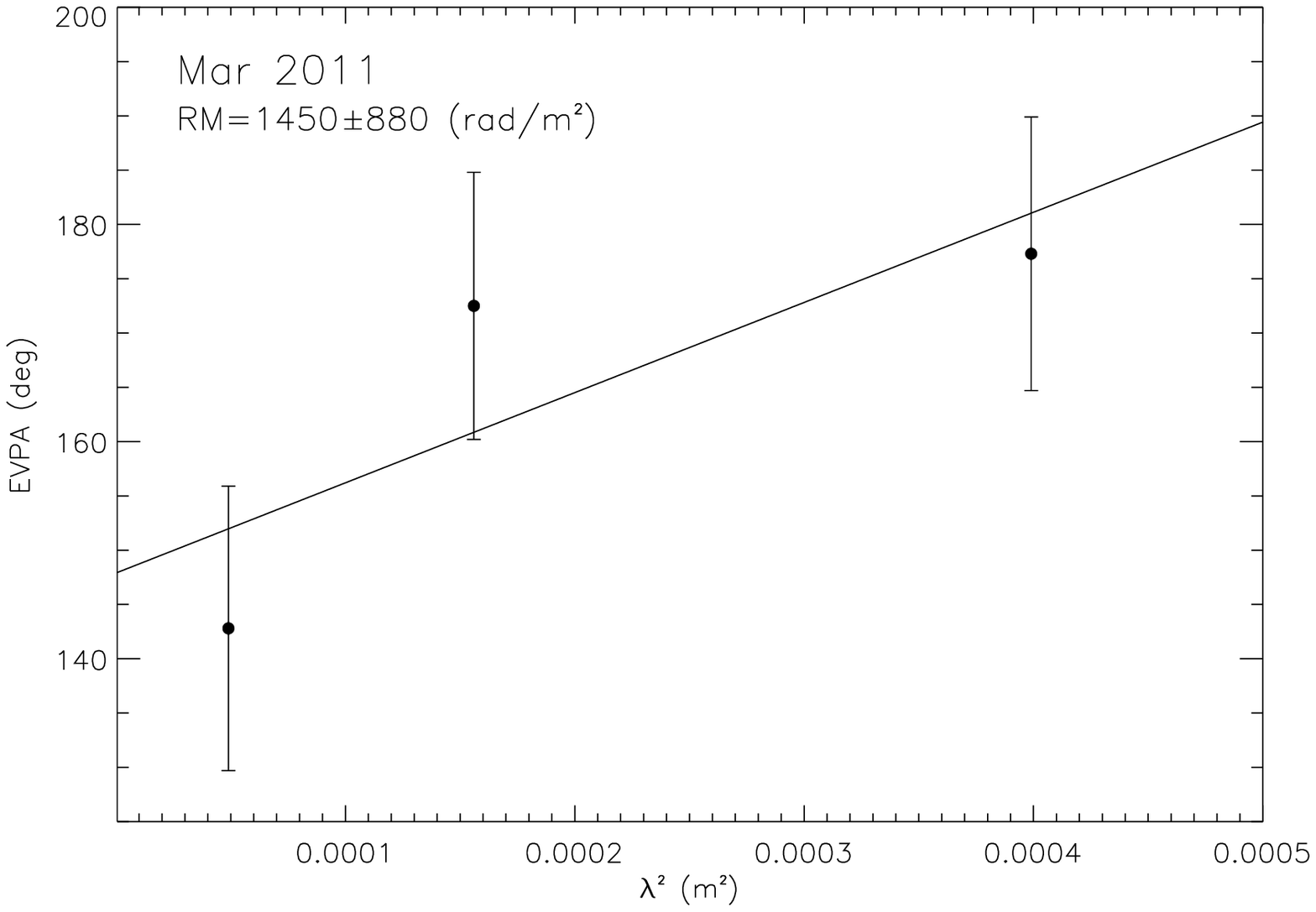} \\      
\includegraphics[width=0.67\columnwidth]{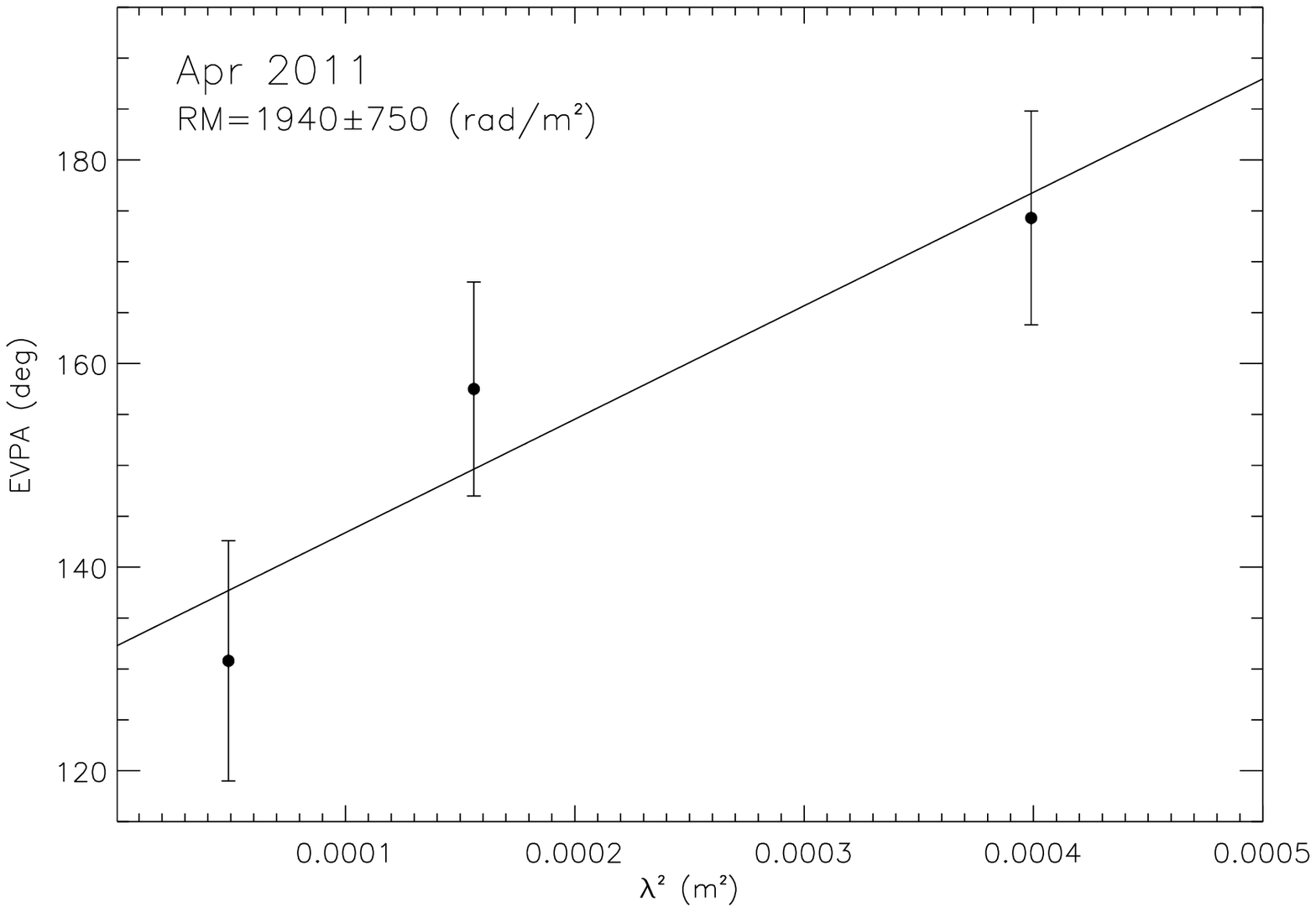}         
\includegraphics[width=0.67\columnwidth]{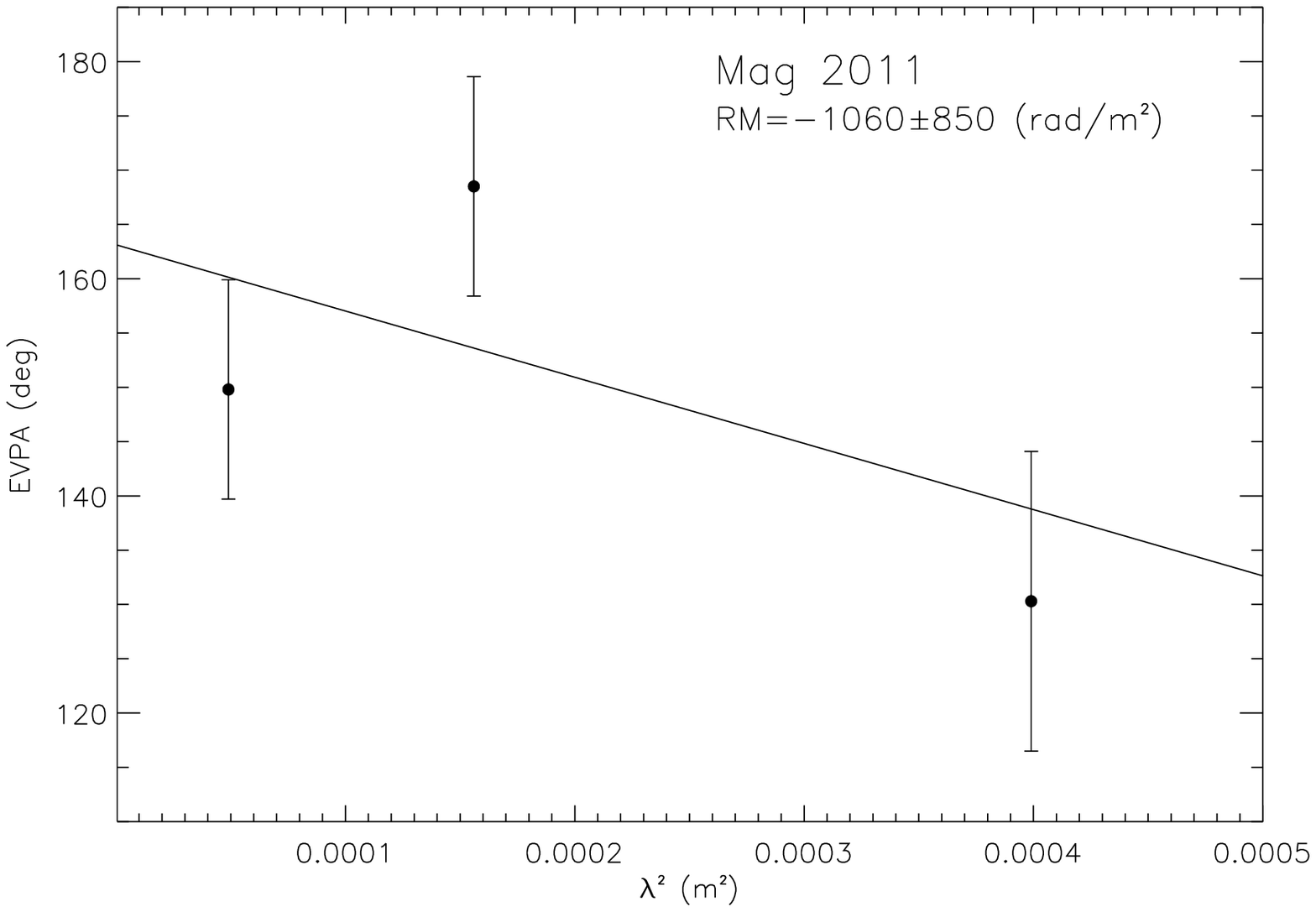}
\includegraphics[width=0.67\columnwidth]{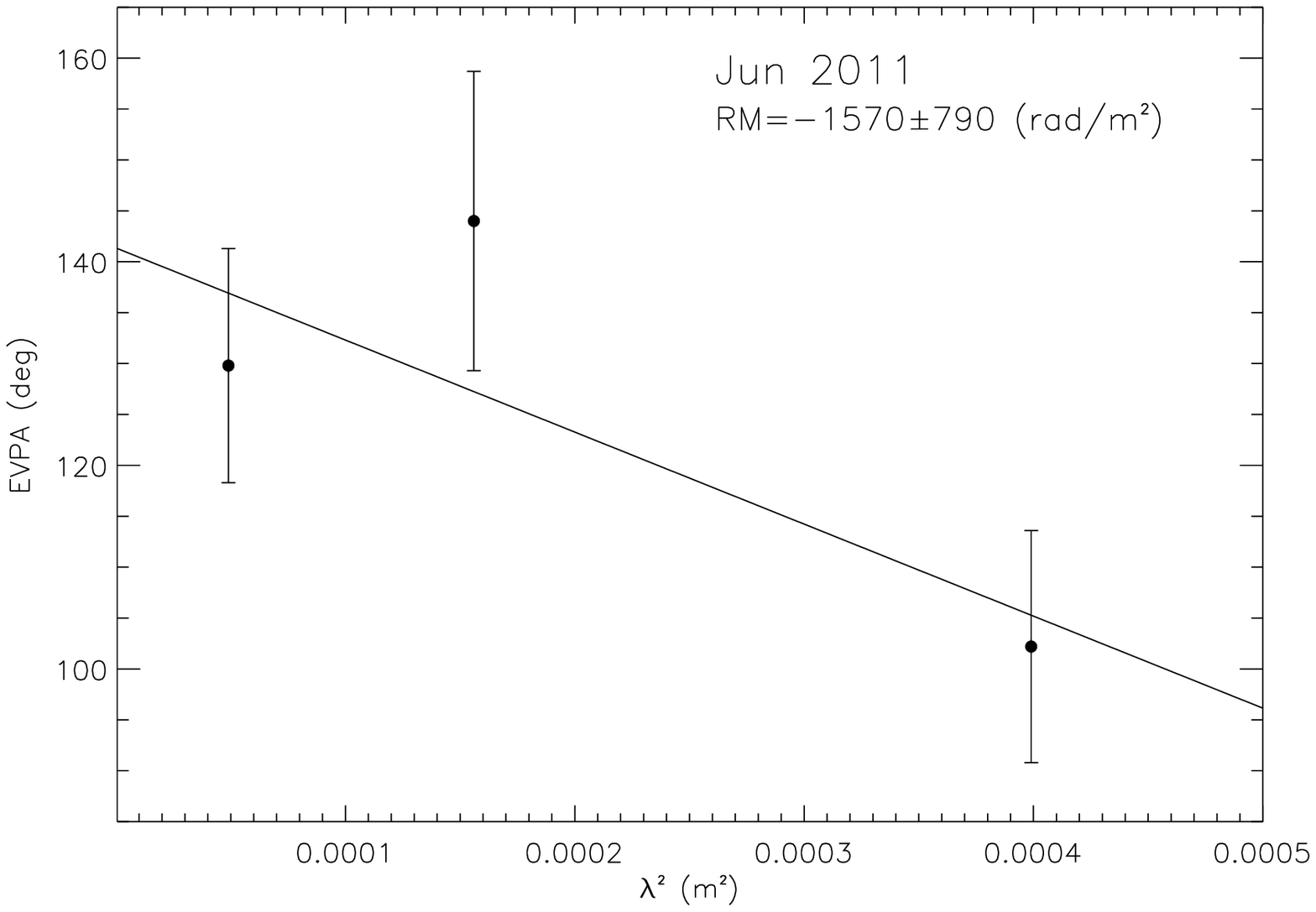} \\
\includegraphics[width=0.67\columnwidth]{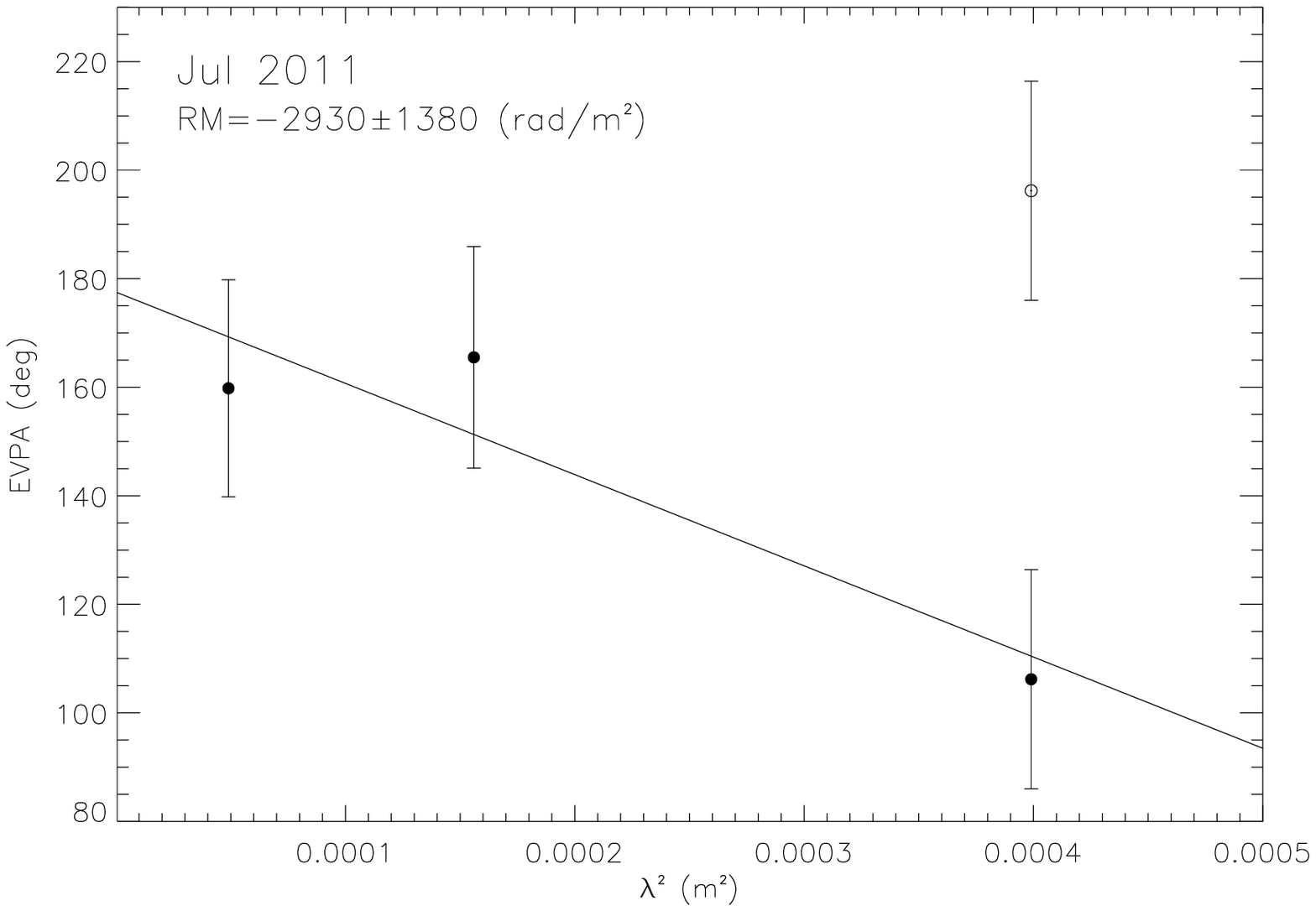}
\includegraphics[width=0.67\columnwidth]{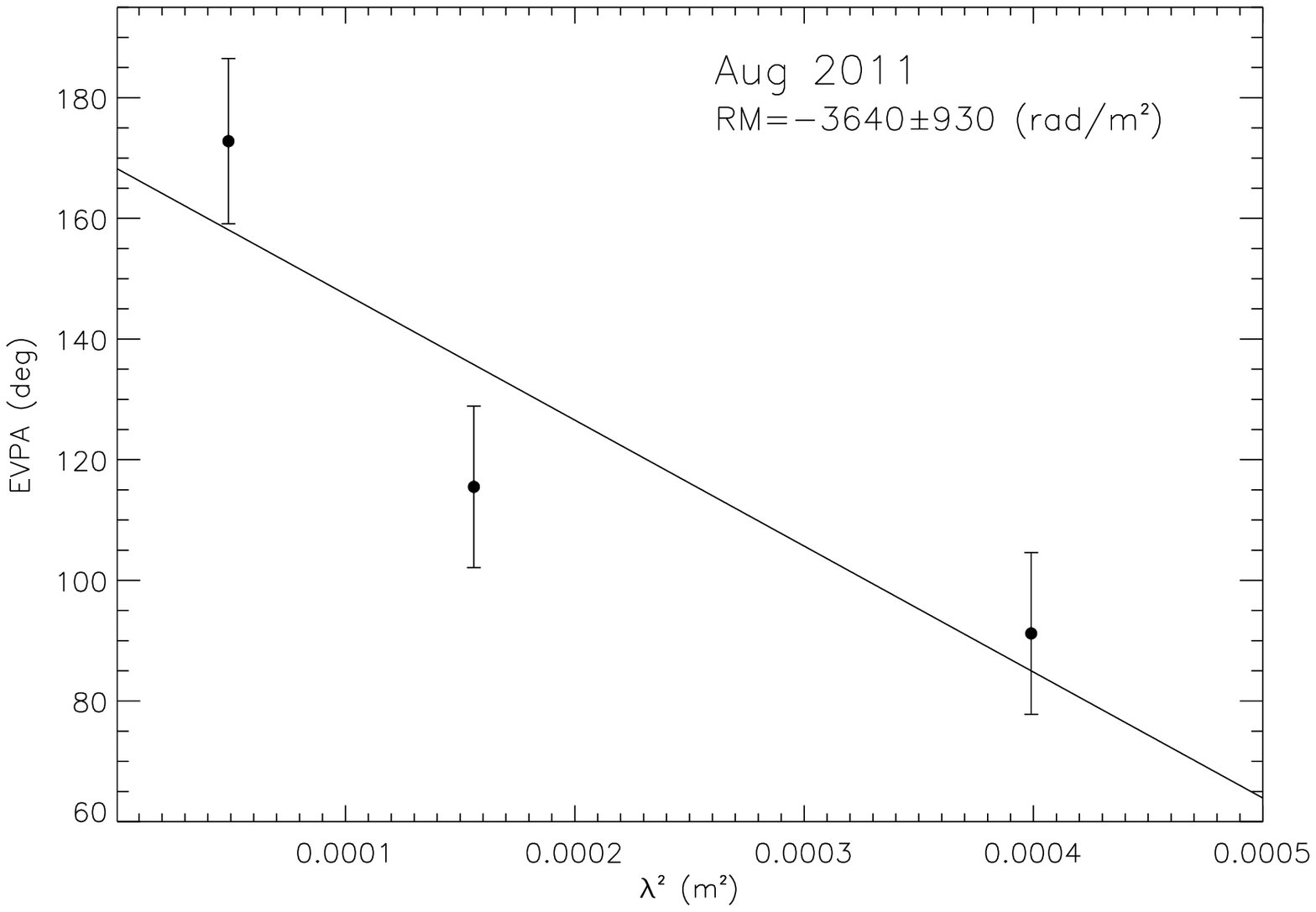}
\includegraphics[width=0.67\columnwidth]{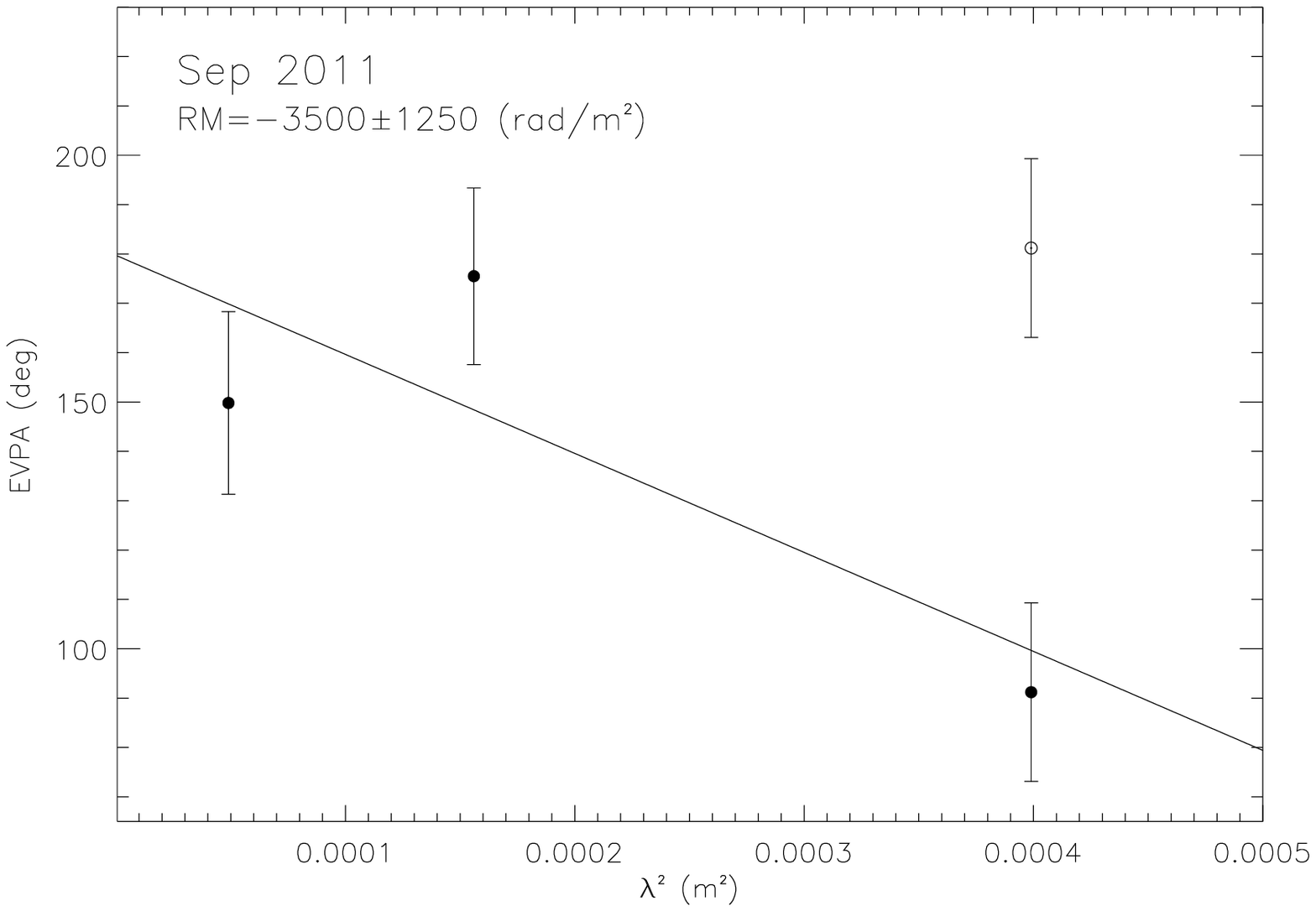} \\
\includegraphics[width=0.67\columnwidth]{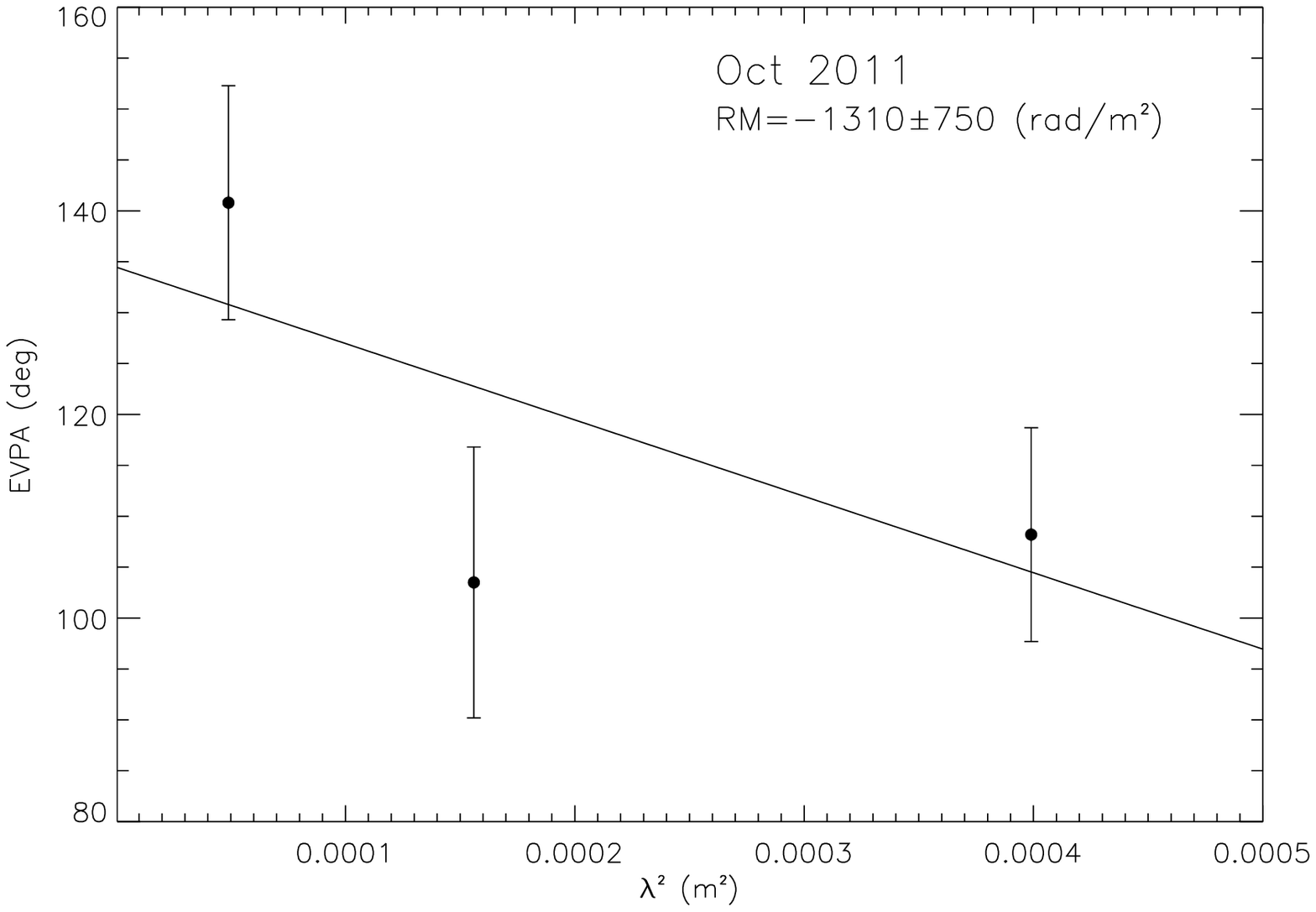}
\includegraphics[width=0.67\columnwidth]{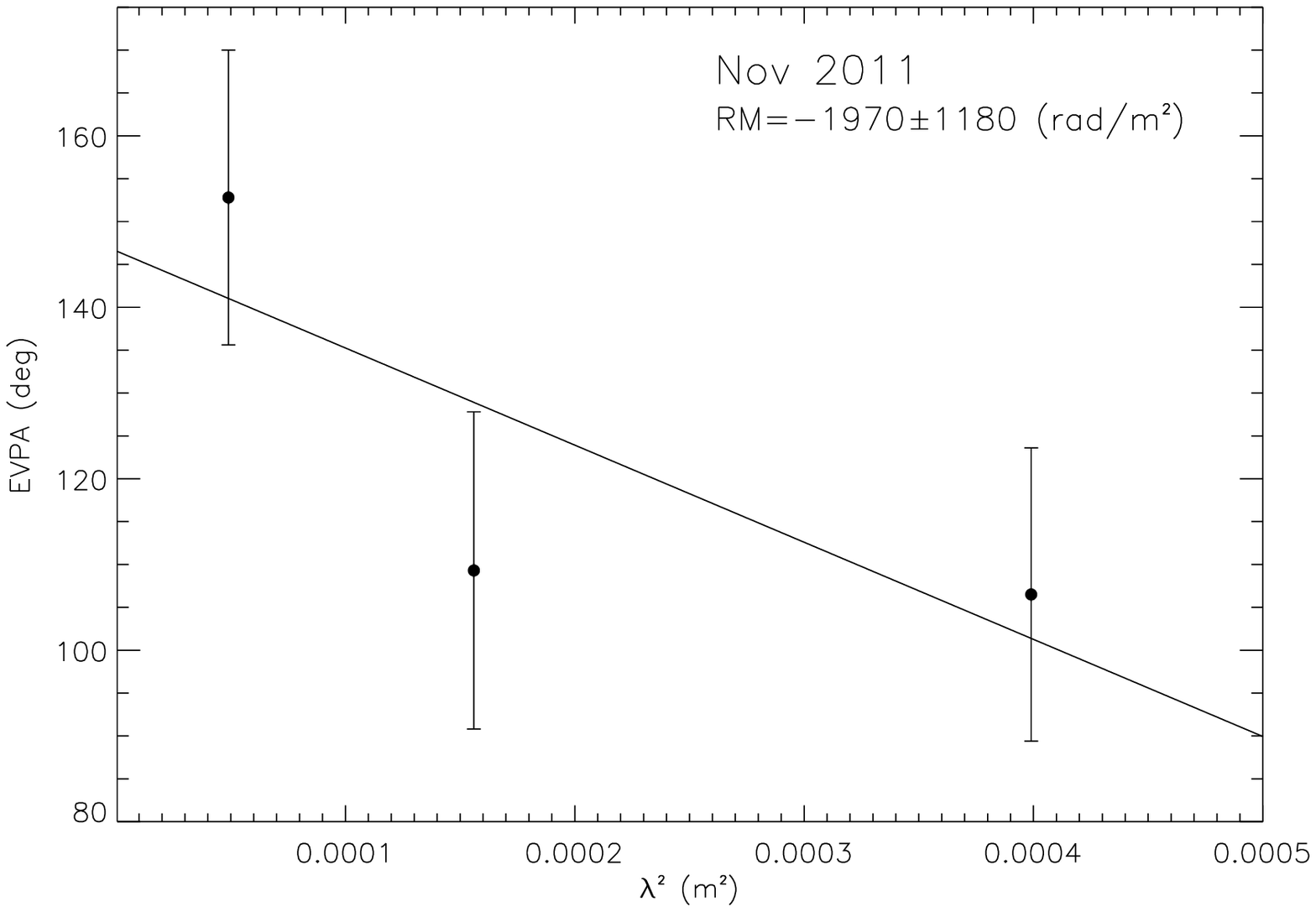}
\includegraphics[width=0.67\columnwidth]{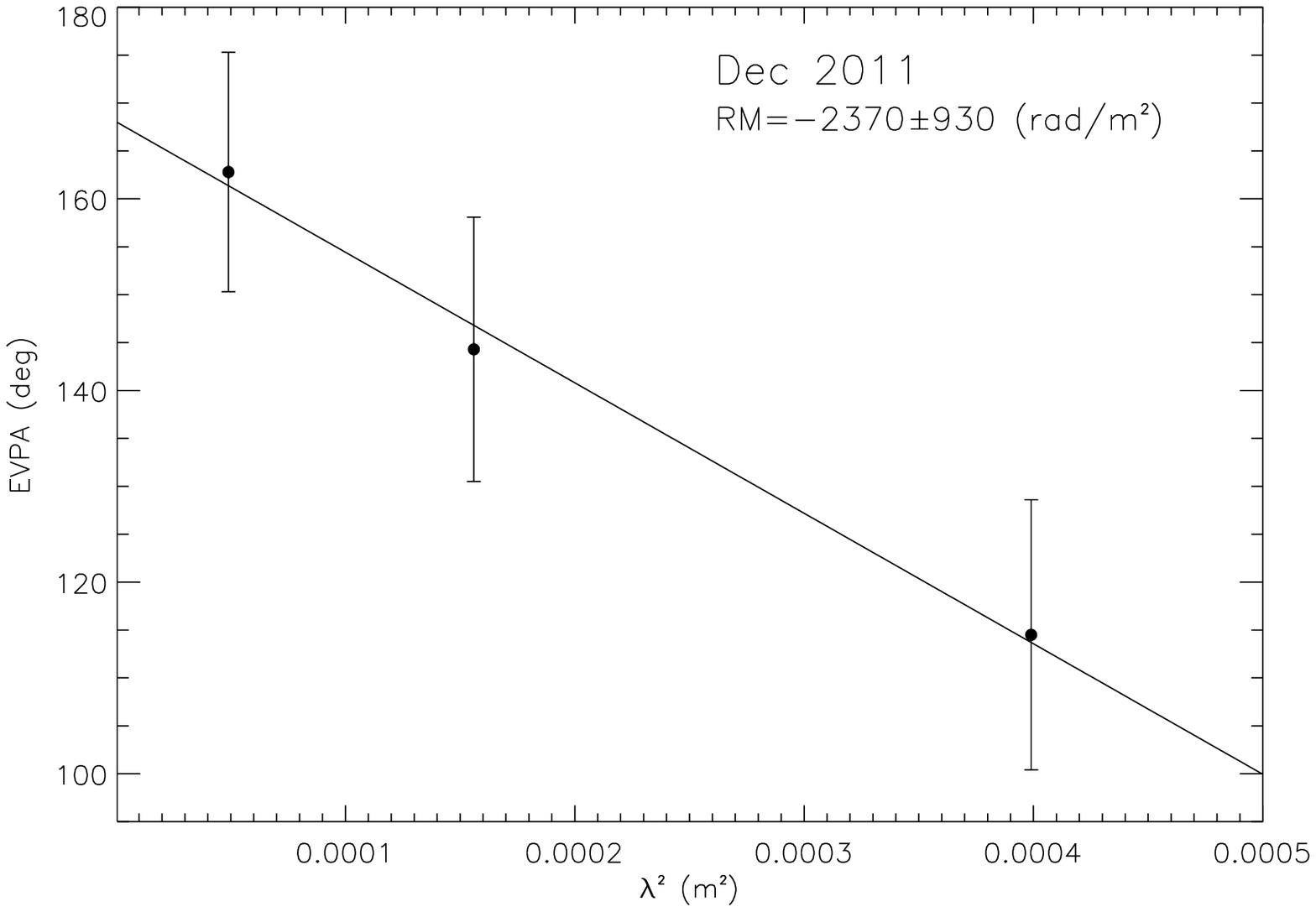} \\
\caption{EVPAs versus $\lambda^2$ linear fits for all the observing epochs. We also show the 15\,GHz EVPAs with empty symbols as measured in the maps, i.e.\ before the rotation of $90^\circ$ applied for July and September to account for opacity effects.} 
\label{rm_core}
\end{figure*}

\section{Results}
\subsection{Images and morphology}
\label{morphology}
In Fig.~\ref{maps} we show a sample of three polarization images of Mrk\,421 at 15\,GHz (upper panel), 24\,GHz (middle panel), and 43\,GHz (lower panel) produced with DIFMAP and IDL\footnote{\url{http://www.exelisvis.com/ProductsServices/IDL.aspx}}. To improve the sensitivity to the extended jet emission, we restored the images with natural weighting. The contours show the total intensity, the overlaid color maps show the linearly polarized intensity. Bars represent the absolute orientation of the EVPAs. 
We note that the position of the peak in the linearly polarized and total intensity emission images do not always coincide. For example, in the first observing epoch at 15\,GHz (upper panel in Fig.~\ref{maps}), the peak is 4.93 mJy/beam, and it is not coincident with the total intensity peak, instead it lies in the jet about $1$ mas from the core region. 
From the total intensity images at all three frequencies, we clearly detect a well-defined and collimated one-sided jet structure, emerging from a compact nuclear region that extends about 4.5 mas (2.7 pc), with a position angle (PA) of about $-35^\circ$. This agrees well with the result of \citet{Giroletti2004a}. 
At 15\,GHz the linearly polarized emission extends for about 1 mas from the core region, allowing us to distinguish core and jet emission. The outer polarized emission is too faint to be detected. The polarized emission in the jet becomes fainter at higher frequencies: at 24\,GHz it is still clearly detected, but at 43\,GHz we only detect polarized emission from the core region; this is probably due to sensitivity limitations. 

In our previous works we represented the total intensity jet structure with Gaussian components. However, this approach is not reliable in the case of polarized emission. We instead estimated the jet polarized flux density as the difference between the total amount and the contribution of the core region. In practice, we first measured the total polarized flux density $P_\mathrm{tot}$ at each epoch by setting a box containing the entire polarized region; we adjusted the size of the box depending on the extension of the polarized emission. We then determined the core contribution $P_\mathrm{core}$ from the value of the polarized flux density at the position of the total intensity peak. Finally, we estimated the jet polarized flux density as $P_\mathrm{jet}=P_\mathrm{tot}-P_\mathrm{core}\times\cos(\chi_\mathrm{core}-\chi_\mathrm{jet})$. We also determined the jet EVPA directly on the image at the location where the jet polarized flux density is highest.

The 43\,GHz images reveal a transverse structure in the inner part of the jet in the form of limb brightening in the polarized emission. This transverse structure is clearly revealed in the  April 2011 polarization image (see lower panel in Fig.~\ref{maps}). During this epoch the polarization emission peak is in the core region and is $\sim8$ mJy/beam, while in the limbs the polarization peak is $\sim2.3$ mJy/beam. The limb-brightening structure is also detected in March and May 2011, but it is less pronounced, while it is absent from all the other epochs.

\begin{figure}
\includegraphics[width=1.0\columnwidth]{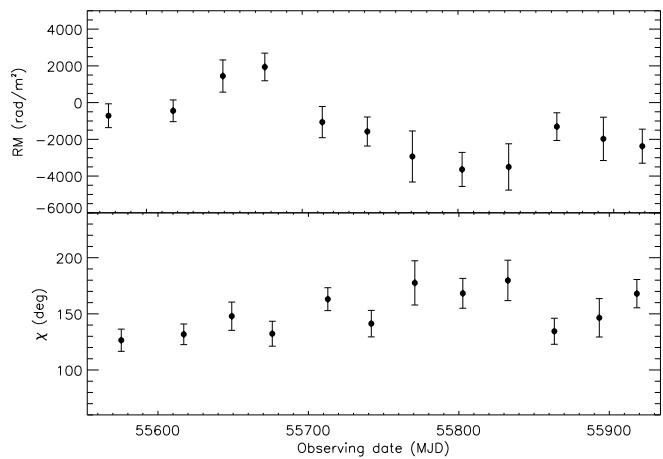} \\
\caption{Upper panel: time evolution for the RM values derived for the core region using 15, 24, and 43\,GHz data. Lower panel: time evolution for the intrinsic values of the polarization angle obtained from the $\lambda^2$ fits.} 
\label{rm}
\end{figure}

\subsection{Radio light curves and evolution of polarization angle} 

In Fig.~\ref{plots_core}, we show the light curves for the core region of Mrk\,421 during 2011 at 15\,GHz (upper frame), 24\,GHz (middle frame), and 43\,GHz (lower frame). For each frame, we show in four panels (from top to bottom) the total intensity and the polarized flux density, the fractional polarization, and the EVPAs. 
The plots for the jet region are shown in Fig.~\ref{plots_jet} at 15\,GHz (upper frame) and 24\,GHz (lower frame). No polarized emission from the jet is apparent in the 43\,GHz images. All of these values are reported in Table~\ref{table_data}.

\begin{figure*}
\includegraphics[width=1.0 \columnwidth]{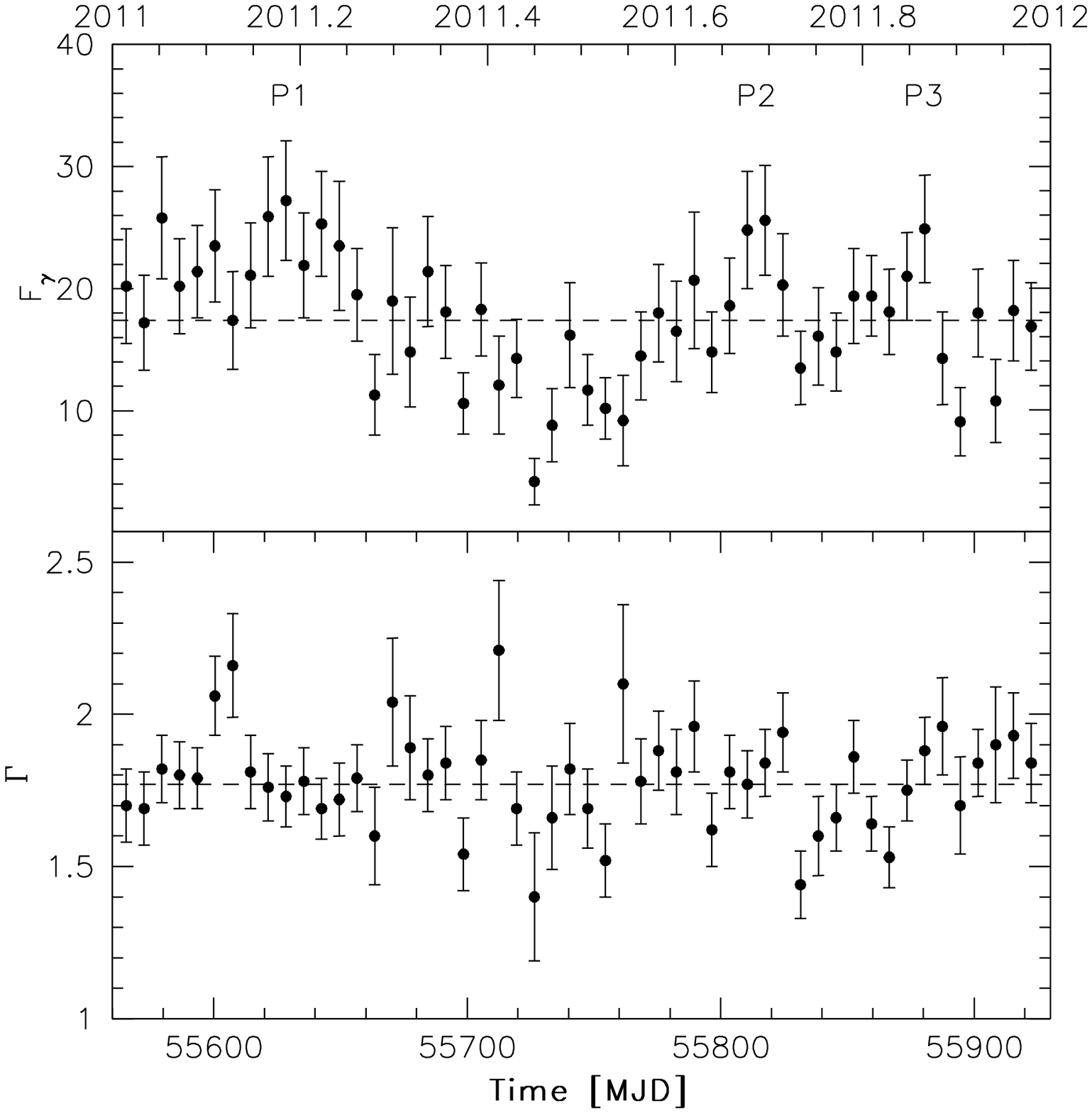}  %
\includegraphics[width=1.0\columnwidth]{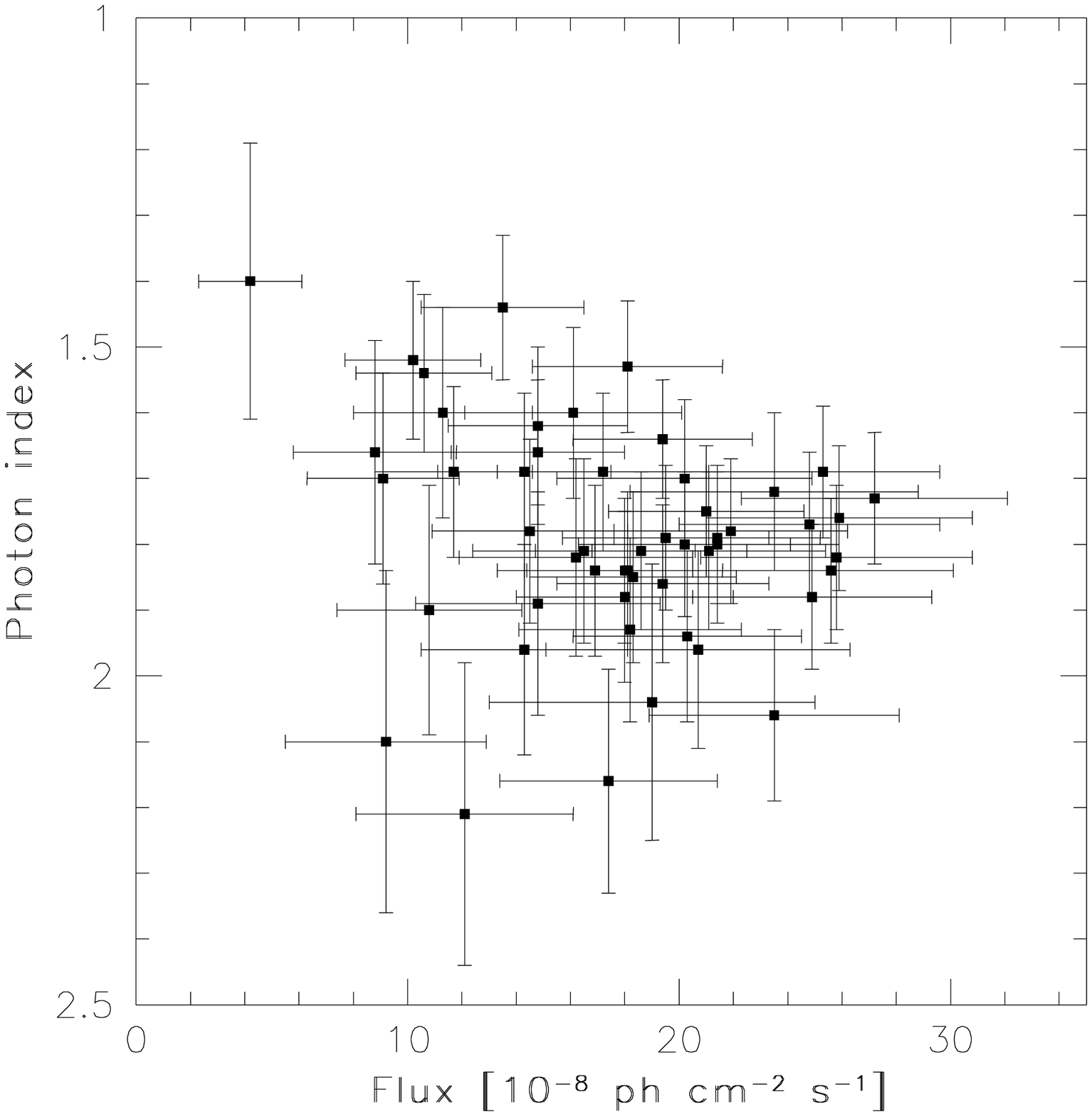}
\caption{0.1-100 GeV flux in units of 10$^{-8}$ ph cm$^{-2}$ s$^{-1}$ (top panel) and the spectral photon index from a power-law fit (bottom panel) for Mrk\,421 for time intervals on one week during 2011 (from MJD 55562 to MJD 55926). The dashed line in both panels represents the mean value. In the right frame we show the scatter plot of the photon index vs the $\gamma$-ray flux.} 
\label{fermi_plot}
\end{figure*}

\subsubsection{Core region}

As we noted in our previous works \citep{Lico2012, Blasi2013}, for all the three frequencies the total intensity light curves peak around the end of February (MJD $\sim$55617), with a decrease until July 28 (MJD 55770) followed by a slight increase in the last part of the year. 

In the core region the linearly polarized emission at 15\,GHz and 24\,GHz is significantly variable ($>3\sigma$ difference between the highest and the lowest value), but we observe the most significant variation at 43\,GHz, where we clearly detect a peak of $12.5$ mJy/beam during the third observing epoch (MJD $\sim$55649). This behavior may be connected with the enhanced activity in the $\gamma$-ray light curve between MJD 55562 and MJD 55660, as described in Sect.~\ref{fermi_sec}. 
 
The core polarization fraction has a mean value of $1\%$, in agreement with other studies of this source \citep[e.g.,][]{Marscher2002, Pollack2003}. At 43\,GHz the polarization percentage is higher, around $2\%$, reaching $\sim4\%$ during the third observing epoch, which is close in time with the enhanced activity at high energy.

In the bottom panel of each frame of Fig.~\ref{plots_core}, we also show the trend of the polarization angle for the core region of Mrk\,421 during 2011 at 15\,GHz (upper frame), 24\,GHz (middle frame), and 43\,GHz (lower frame). The EVPAs vary. For most of the year, they show values varying between $110^\circ$ and $150^\circ$. In between some pairs of epochs, in particular at 15\,GHz, they change by about $90^\circ$, which has no clear connection with the EVPA variation at 24 and 43\,GHz or with the polarized flux density trend. 

\begin{table}
\caption{Rotation measure and intrinsic polarization angle values for the core region.}
\label{rm_table}     
\centering    
\tiny                 
\begin{tabular}{c c c c c c}        
\hline\hline
Epoch & MJD & RM\tablefootmark{a} & $\sigma_{RM}$\tablefootmark{b} & $\chi$\tablefootmark{c} & $\sigma_{\chi}$\tablefootmark{d} \\
year/month/day & & (rad\,m$^{-2}$) & (rad\,m$^{-2}$) & ($^\circ$) & ($^\circ$) \\
\hline
2011/01/14 & 55576 & $-710$  & 650  & 127 & 10\\
2011/02/25 & 55617 & $-450$  & 590  & 132 & 9\\
2011/03/29 & 55649 & 1450  & 880  & 148 & 13\\
2011/04/25 & 55676 & 1940  & 750  & 132 & 11\\
2011/05/31 & 55713 & $-1060$ & 850  & 163 & 10\\
2011/06/29 & 55742 & $-1570$ & 790  & 141 & 12\\
2011/07/28 & 55771 & $-2930$ & 1380 & 178 & 20\\
2011/08/29 & 55803 & $-3640$ & 930  & 168 & 13\\
2011/09/28 & 55833 & $-3500$ & 1250 & 180 & 18\\
2011/10/29 & 55863 & $-1310$ & 750  & 135 & 12\\
2011/11/28 & 55893 & $-1970$ & 1180 & 147 & 17\\
2011/12/23 & 55918 & $-2370$ & 930  & 168 & 13\\
\hline 
\hline
\end{tabular}
\tablefoot{
\begin{tiny}
\newline
\tablefoottext{a}{Rotation measure in rad\,m$^{-2}$.}\\
\tablefoottext{b}{Estimated error for the rotation measure.}\\
\tablefoottext{c}{Intrinsic polarization angle in deg.}\\
\tablefoottext{d}{Estimated error for the intrinsic polarization angle.}\\
\end{tiny}
}

\end{table}

\subsubsection{Jet region}

The light curves for the jet region, extending to about 1 mas from the core at 15 and 24\,GHz, are shown in Fig.~\ref{plots_jet}.  The situation is very different in the jet from the core region. The total intensity flux density for the extended region does not show any significant variation. 
The polarized flux density shows some variability, reaching a peak during the fourth epoch at 15 and 24\,GHz.
The polarization percentage in the jet region is about $16\%$; this higher degree of polarization with distance from the core seems to be a common feature in blazars \citep{Lister2001}. Finally, the EVPAs are also quite stable, fluctuating weakly around a value of $60^\circ$, i.e., roughly perpendicular to the jet position angle ($\sim-35^{\circ}$).

\subsection{Faraday rotation analysis}
\label{faraday}
A polarized wave propagating through a magnetized plasma is affected by Faraday rotation. As a consequence, the observed polarization angle ($\chi_\mathrm{obs}$) appears rotated with respect to its intrinsic value ($\chi_\mathrm{int}$). This effect is described by a linear relationship between $\chi_\mathrm{obs}$ and the observing wavelength squared ($\lambda^2$):

\begin{equation}
\chi_\mathrm{obs}=\chi_\mathrm{int} + RM\times \lambda^2,
\end{equation}
where RM represents the rotation measure, a quantity related to the electron density $n_e$ (cm$^{-3}$), the parallel to the line of sight, aberrated by relativistic motion, component of the magnetic field $\textbf{B}_{\parallel}$ (milligauss), and the path length $dl$ (parsecs):

\begin{equation}
RM = 812 \int n_e \textbf{B}_{\parallel} \cdot dl \ \ \ [\mbox{rad \ m}^{-2}].
\end{equation}

For the core region, where EVPA values at 15, 24, and 43\,GHz for each observing epoch are available (see Table~\ref{table_data}), we performed linear fits of EVPAs versus $\lambda^2$, obtaining the RM and $\chi_\mathrm{int}$  values and uncertainties reported in Table~\ref{rm_table}. Since the two flips observed at 15 GHz in July and September strongly suggest optically thin-thick transitions, we carried out RM fits using EVPA values rotated by $90^\circ$ at 15 GHz in these epochs. All of these linear fits are reported in Fig.~\ref{rm_core}; in some cases they show significant scatter about a linear trend (e.g., in February 2011), and in other cases they agree well with a linear behavior (e.g., in December 2011).
The time evolution of RM values and of the intrinsic EVPA values are reported in the upper and lower panels of Fig.~\ref{rm}, respectively.

The RM values are distributed across a wide range of values, spanning $(-3640 \pm 930)$ to $(+1940\pm750)$ rad\,m$^{-2}$. However, the uncertainties are often very large, with many values being consistent with 0 within $1\sigma$ or $2\sigma$. It is thus difficult to provide accurate values and even more difficult to claim significant variability during the year. 
For the intrinsic values of the polarization angle, the results {\bf tend to} reflect the roughly stable behavior observed for the 43\,GHz EVPAs, with a value of about 150$^\circ$, that is, roughly parallel to the jet axis (see Fig.~\ref{rm}). However, there is some residual, mildly significant variability ($F_{var}$ is $0.10 \pm 0.04$).


\subsection{$\gamma$-ray light curves from {\em Fermi} data}
\label{fermi_sec}
After integrating over the period 2011 January 1 - December 31, we obtain a fit  with a power-law model in the 0.1-100 GeV energy range that results in TS = 8728 ($\sim 93 \, \sigma$), with an average flux of ($17.4 \pm 0.5) \times10^{-8}$ ph cm$^{-2}$ s$^{-1}$, a photon index of $\Gamma = 1.77 \pm 0.02$, and an apparent isotropic $\gamma$-ray luminosity of $7.5 \times10^{44}$ erg s$^{-1}$, which is fully compatible with the values obtained over the first two years of {\em Fermi} operation \citep{Nolan2012}. In Fig.~\ref{fermi_plot} we show the $\gamma$-ray flux using time bins of one week (top panel) and the photon index variation (bottom panel) over the entire observing period. For each time bin, the spectral parameters for Mrk\,421 and for all the sources within a radius of 10$^{\circ}$ were left free to vary. 

In the $\gamma$-ray light curve we identify three peaks: a main peak (P1) in the first observing period (2011 March 5-11, MJD 55625-55631) with a subsequent decrease to the lowest flux level in 2011 June, followed by two other peaks in the final observing period (P2 in 2011 September 3-16, MJD 55807-55820; P3 in 2011 November 12-18, MJD 55877-55883). 

The daily peak flux during P1 is $(38\pm11)\times10^{-8}$ ph cm$^{-2}$ s$^{-1}$, observed on 2011 March 7, corresponding to an apparent isotropic $\gamma$-ray luminosity of $1.6\times10^{45}$ erg s$^{-1}$.
The daily peak flux during P2 and P3 is $(37\pm12)\times10^{-8}$ ph cm$^{-2}$ s$^{-1}$ and $(30\pm9)\times10^{-8}$ ph cm$^{-2}$ s$^{-1}$, observed on September 8 and November 13, respectively.

Spectral hardening during flares has been seen in some blazars, FSRQs in particular \citep[e.g., PKS 1510$-$089;][]{D'Ammando2011}. This sort of behavior is not without precedent in BL Lacs, but it is rare \citep[e.g.,][]{Gasparrini2011,Raiteri2013}. Consistent with the trend for BL Lacs in general, and for Mrk 421 in particular (e.g., Abdo et al 2011), we do not detect any spectral hardening in Mrk 421 during the periods of enhanced $\gamma$-ray activity, and no obvious relation is observed between the photon index versus the $\gamma$-ray flux (see Fig.~\ref{fermi_plot}). Furthermore, the spectral index is generally compatible with the average, ranging from $1.4$ to $2.2$.

\subsection{Correlation between radio and $\gamma$-ray data}
\label{corr_sec}
The comparison of the 15, 24, and 43 GHz radio core and $\gamma$-ray light curves suggests similar trends. We calculated the correlation coefficient between the radio core flux density at each epoch and frequency and the $\gamma$-ray flux during the weekly period containing the radio observations. We report these results in Table~\ref{t.coeff}. All of the radio data sets show a strong mutual correlation, with coefficients in the range $0.88-0.96$. The correlation coefficients between radio and $\gamma$-ray data are lower, in the range $0.42-0.46$; the strongest correlation is found with the data at 43\,GHz, where the largest number of data points is available. 

\begin{figure}
\includegraphics[bb=50 0 504 360,width=1.0\columnwidth ,clip]{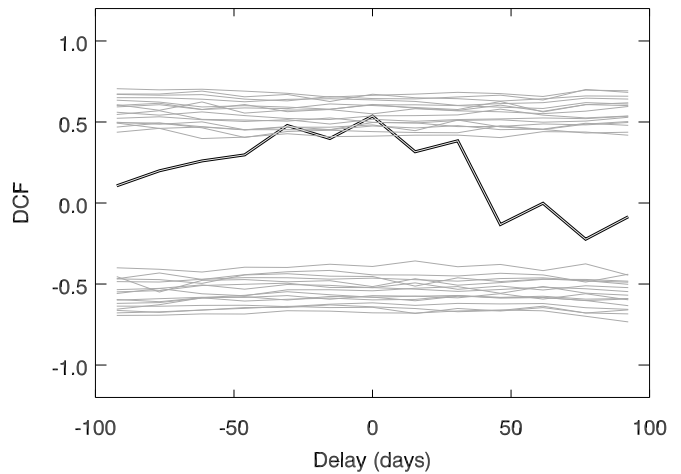}
\caption{Discrete cross-correlation function between the $\gamma$-ray and 43 GHz radio light curves (black curve). The gray curves represent the 99.7\% confidence limits relative to stochastic variability, obtained from the combination of different power spectral density slopes. See Sect. \ref{corr_sec} for more details.} 
\label{DCF}
\end{figure}

To assess the significance of this correlation and to determine a possible time lag, we calculated the discrete cross-correlation function (DCF) between the radio core flux density at 43\,GHz and the $\gamma$-ray flux.
The results of the correlation analysis are shown in Fig.~\ref{DCF}. \\
To compute the DCF, we used the algorithm developed by \citet{Edelson1988}, and to determine the significance of the correlation, we followed the approach recommended by \citet{Chatterjee2008} and \citet{Max-Moerbeck2010}. As recommended by \citet{Timmer1995}, we simulated 3000 light curves with the same mean and standard deviation as the observed light curves.
The power spectral density (PSD), corresponding to the power in the variability of emission as a function of timescale, is represented by a power-law $\mbox{PSD} \propto f^{-\beta}$, where $f$ is the inverse of the timescale.
For each set of simulations, $\beta$ varies from 1 to 2.5 in steps of 0.5. The curves obtained from different combinations of different PSD slopes, with a confidence level > 99.7\%, are shown in gray in Fig.~\ref{DCF}.

We investigated the delay over a range of $\pm100$ days, with a bin of 15 days. The highest value for the correlation (0.54) is found for zero delay. As is clear from Fig.~\ref{DCF}, this peak does not have a high significance level
for all combinations of the different PSD slopes used here. It has a significance level $> 99.7\%$ for the combination of $\beta$ lower than 1.5 for the $\gamma$-ray PSD and $\beta$ ranging from 1 to 2.5 for the radio PSD. This agrees with the fact that we observe the strongest variability on shorter timescales for the $\gamma$-ray emission (flatter $\beta$) and on longer timescales for the radio emission (steeper $\beta$).

\section{Discussion}

For many decades, Mrk\,421 has been a target of regular monitoring at radio frequencies, showing only moderate variability \citep{aller1999,Venturi2001,Fan2007,Nieppola2009}. Multi-wavelength campaigns have failed to reveal significant correlations between variability at high (or very high) energy and radio wavelengths \citep{Abdo2011,Acciari2011}. The variability detected within the present campaign and its possible connection to the $\gamma$-ray light curve are therefore of great interest, especially considering that they foreshadowed the dramatic and unprecedented radio and $\gamma$-ray flares in 2012 \citep{D'Ammando2012,Richards2013}.

\subsection{Possible radio and $\gamma$-ray connection}
\begin{figure}
\includegraphics[width=1.0\columnwidth]{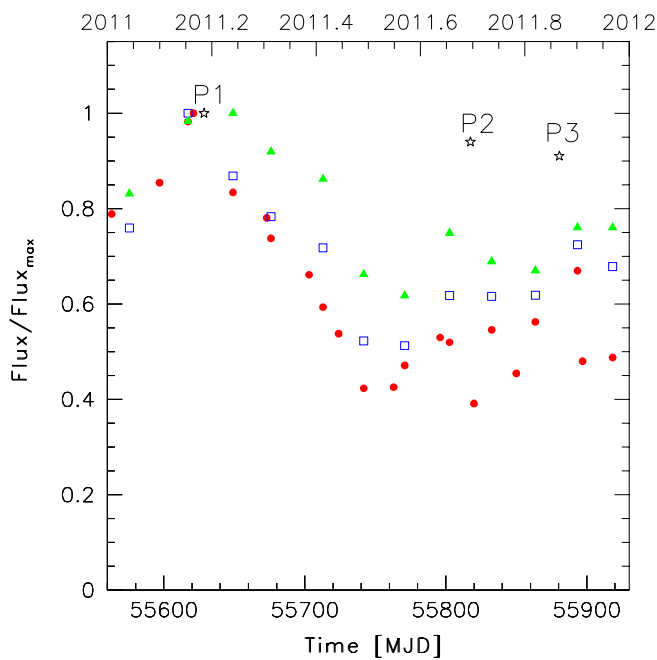}
\caption{Radio light curves normalized to the peak value. Triangles, squares, and circles represent 15, 24, and 43\,GHz data, respectively. All error bars (not shown) are about 0.1 mJy.  The stars (labeled with P1, P2 and P3) represent the three main peaks in the $\gamma$-ray light curve normalized to the highest value.} 
\label{norm_light}
\end{figure}

We find significant variability in the first months of 2011 at all three observing frequencies (15, 24, and 43\,GHz) in the total intensity emission ($>3\sigma$ difference between the highest and lowest value). During the same observing period, enhanced activity also occurred at high energies. In Fig.~\ref{norm_light} we show the radio and $\gamma$-ray light curves normalized to the peak value. 

In the radio band, the increase of the flux density is stronger at 43\,GHz, where the fractional variability amplitude $F_{var}$ \citep{Edelson2002, Vaughan2003} is $0.24 \pm 0.06$, while at longer wavelengths it corresponds to $0.13 \pm 0.04$ and $0.18 \pm 0.05$ at 15\,GHz and 24\,GHz. In the $\gamma$-ray light curve $F_{var}$ is $0.17 \pm 0.04$, on a weekly timescale. In both radio and $\gamma$-ray energy bands, the peak value is reached close in time, although the sampling interval of our observations does not allow us to determine the date of the radio burst with a better accuracy than a few weeks. 

\begin{table}
\caption{Pearson correlation coefficients between radio flux density and $\gamma$-ray photon flux.}
\label{t.coeff}     
\centering    
\begin{tabular}{c c}        
\hline\hline
Data pairs & $r$ \\
\hline                        
$r_{15-24}$ & 0.93\\
$r_{15-43}$ & 0.88 \\
$r_{24-43}$ & 0.96 \\
$r_{15-\gamma}$ & 0.44\\
$r_{24-\gamma}$ & 0.42\\
$r_{43-\gamma}$ & 0.46\\
\hline 
\hline
\end{tabular}
\end{table}

The main peak in the radio light curve (around MJD 55621) occurs close in time with the first $\gamma$-ray peak (MJD 55627); no clear radio counterpart was observed for the second and third $\gamma$-ray enhanced activity episodes. Overall, we find a good correlation between the low-frequency and high-energy emission, as shown by the high and statistically significant value of the correlation coefficient. This could indicate a co-location of the radio and $\gamma$-ray emission regions, and a size as compact as $c\, \Delta t\, \delta \sim 5.3\times10^{16}\times \delta$ cm, assuming $\Delta t \sim 3$ weeks.

On the other hand, the study of the possible delay between radio and $\gamma$-ray light curves, based on the DCF, did not provide significant constraints on the lag between the radio and $\gamma$-ray light curves. In fact, much longer data trains are necessary to reach highly significant lag values \citep[for a detailed discussion of the significance of radio-gamma light curve correlation and lags, see][]{Max-Moerbeck2013}. We furthermore note that in a recent study of the radio and $\gamma$-ray correlated variability in {\it Fermi} bright blazars based on 3.5 years of dense monitoring, \citet{Fuhrmann2014} did not find significant correlation between the 7mm and the $\gamma$-ray data for Mrk\,421.

The proximity between the radio and $\gamma$-ray enhanced activity at the beginning of 2011 differs from what was observed in 2012, where the $\gamma$-ray flare \citep{D'Ammando2012} led the radio burst by about 40 days at 15\,GHz \citep{Richards2013}. However, the enhanced activities observed in 2011 and 2012 are different. The latter is characterized by a $\gamma$-ray apparent luminosity of $6.7\times 10^{45}$ erg s$^{-1}$, that is, about four times more powerful than the former, with simultaneous tentative detection in the TeV energy band \citep{Bartoli2012}. The different behavior of the two flares may suggest different regions for the $\gamma$-ray activity: upstream of the radio emission in the 2012 flare, while downstream along the jet where the emission is not opaque at the radio frequencies in the 2011 episode.

Interestingly, after the main peak in February 2011, which was observed both at radio and high energies, the polarized flux density at 43\,GHz increases rapidly (Fig.~\ref{plots_core}) and peaks on 2011 March 29 (MJD 55649). Simultaneously the polarization percentage increases at 43\,GHz and reaches a peak value of $3.6\%$, followed by a gradual decrease to the mean value ($\sim 1\%$). A similar behavior was observed by \citet{Piner2005} after an enhanced activity at TeV energies, when the polarization percentage of the core reached $\sim5\%$ at 22\,GHz. This behavior is less pronounced in our data at 15 and 24\,GHz. 

In the jet region the polarized flux density and the fractional polarization reach a peak on 2011 April 25 (MJD 55675) at 15 and 24\,GHz. These peaks occur just after the enhanced activity observed in the core region in total intensity emission. This behavior could be associated with a propagating disturbance, but no new components were detected in the total intensity images.

\citet{Blasi2013} investigated the behavior in the optical regime during 2011, by considering optical data provided by the Steward Observatory of the University of Arizona. We found a peak in the V magnitude occurring during the same period. The optical polarization increased from $\sim1\%$ to $\sim7\%$ and the optical polarization angle reduced its scatter. This may indicate that the magnetic field underwent a transition from a tangled state to a more regular one, in correspondence to the observed enhanced activity.

\subsection{EVPA variation and magnetic field topology}
To fully understand the possible relation between the enhanced activity and the physical changes occurring in the source, it is important to investigate the EVPA trend in the different regions of the source. So far, no obvious relation has been found between EVPA variation and high-energy activity. 

In the jet region, extending about 1 mas from the VLBI core, the EVPAs maintain a stable value of about $60^\circ$ during the entire observing period, roughly perpendicular to the inner jet position angle, which is $\sim-35^{\circ}$. This implies that the magnetic field is parallel to the jet axis, which is a rare behavior in BL Lac objects. However, we note that previous VLBI observations of Mrk\,421 had indicated this peculiar magnetic field configuration in its parsec scale jet \citep{Piner2005}. This configuration could be caused either by velocity shear across the jet or by a helical magnetic field with a pitch angle smaller than $45^\circ$ \citep{Wardle2013}. At 24\,GHz, we observe an EVPA rotation of about $60^\circ$ at the end of 2011 (MJD 55893). This behavior may be associated with the propagation of a shock, but no new component and no significant motion were detected in total intensity images. We note that the nondetection of the jet polarized emission at the following epoch (MJD 55918) results in a lack of the EVPA, which does not allow us to confirm the possible EVPA rotation.

In the core region, the EVPA trend is more complex. At 43\,GHz, where the Faraday rotation effects are weaker, the EVPAs are roughly stable in a range of values between $\sim100^\circ$ and $\sim150^\circ$, that is, roughly parallel to the jet axis. This trend is then reflected in the intrinsic values of the polarization angle, obtained from the $\lambda^2$ fits (see the lower panel in Fig.~\ref{rm}). Assuming that this emission region is optically thin at 43\,GHz, the EVPA configuration reflects a magnetic field nearly orthogonal to the jet axis, which is typical of the emission from a transverse shock and different from what is observed in the outer jet. A similar behavior was reported in \citet{Piner2005}: they found parallel EVPAs both in the core and in the innermost jet component (at $\sim$0.3 mas from the core). The extent of the validity of the optically thin emission assumption can be debated. However, the average core spectral index between 15 and 43\,GHz in our data is $\alpha = 0.20\pm 0.13 $ \citep[see also][]{Blasi2013}, suggesting that at 43\,GHz the emission region is predominantly optically thin.

Considering the lower frequencies, in particular at 15\,GHz, we find at least two clear flips of the core EVPAs by about $90^\circ$ (from parallel to perpendicular to the jet, in July and September, with a smaller change also in March). A similar behavior for the EVPAs in the VLBI core region of Mrk\,421 was observed also by \citet{Charlot2006}. They measured a rotation of $\sim60^\circ$ during a three-month period at 15 and 22\,GHz. This behavior is the typical signature of a transition between an optically thin and thick regime, in which the EVPA swings from perpendicular to parallel to the magnetic field direction. Indeed, both in July and September the spectral index flattens ($\Delta \alpha_\mathrm{Jul-Jun}^{15-43}=-0.17$ and $\Delta \alpha_\mathrm{Sep-Aug}^{15-43}=-0.13$). To account for these variations solely by an opacity effect, we adopted in this discussion the core 15 GHz EVPAs rotated by $90^\circ$ for the July and September data, as we already did in determining the RM fits in Sect.~\ref{faraday}.

To explain the observed variability at longer wavelengths, a possible physical scenario might be an association with opacity effects and variable Faraday rotation. The RM variability might be related to changes in the accretion rate, as suggested also by the similar trends found for the RM and the core flux density evolution; for viable explanations of the RM sign change in relativistic jets, see also \citet{OSullivan2009} and \citet{Gomez2011}.

Finally, residual  variability ($F_{var}$ is $0.10 \pm 0.04$) is present in the intrinsic (Faraday de-rotated) EVPAs. This intrinsic variability can be the consequence of a blend of variable cross-polarized subcomponents within the beam, whose relative contributions to the total polarization properties vary as a function of time. This structure explains both the variations of the intrinsic EVPA values, which are integrated over the VLBI core region, and the low polarization of the core, resulting from significant cancellation from subcomponents with different EVPAs.

The cross-polarized subcomponents can be explained by two different physical scenarios: (1) the jet base might be turbulent, so that the subcomponents would be turbulent cells with random field direction; (2) the core magnetic field might be perpendicular to the jet, but with much of the emission optically thick at 15\,GHz, giving rise to EVPAs parallel to the jet in the optically thin part and EVPAs perpendicular to the jet in the optically thick part. 
In either scenario the intrinsic EVPA variations are the direct consequence of changes in the relative weight of one component with respect to the other(s). The second scenario seems to be favored by the observed behavior at 43\,GHz (EVPAs stable in the range 100$^\circ$-150$^\circ$).

The relevance of subcomponent blending can be directly seen in the April data (MJD 55675): the 43\,GHz polarization image clearly shows a transverse EVPA distribution and a limb-brightened structure in the polarized emission (see the lower panel in Fig.~\ref{maps}) in the inner part of the jet. We argue that velocity gradients in this region might cause Kelvin-Helmholtz instabilities that would heat the boundary layer, causing the limb brightening. A similar transverse structure for this source was observed by \citet{Piner2010} at 22\,GHz. These authors noted that the EVPAs were almost parallel along the jet axis and became orthogonal toward the jet edges. This is similar to what we observe in the EVPAs in the western limb. This may reflect a spine plus layer polarization structure and seems to be a common feature in TeV blazars, such as Mrk\,501 \citep[e.g.,][]{Giroletti2008, Piner2010}. 

The observed transverse EVPA rotation implies a complex topology for the magnetic field. For example, \citet{Lyutikov2005} argued that such a transverse EVPA distribution may be consistent with a large scale helical magnetic field in a resolved cylindrical jet.
This transverse polarization structure would reflect the intrinsic magnetic field geometry, and not the propagation of a shock arising from the core and interacting with the surrounding medium \citep{Lyutikov2005, Gabuzda2004}.
Finally, we observe that, contrary to what was found by \citet{Piner2010}, no increase of the fractional polarization towards the jet edges is observed in our data. This may be due to the high activity state of Mrk\,421, causing an enhancement of the polarized emission from the core region.

\section{Summary and conclusions}
We presented a detailed analysis of new VLBA polarization observations of Mrk\,421 at 15, 24, and 43\,GHz during 2011. 

During this observing period the source showed significant radio flux density variability. In particular, in the first part of 2011 we detected a prominent peak in the radio light curves and an enhanced $\gamma$-ray activity occurring close in time. A good correlation was found between the radio core and the $\gamma$-ray light curves. Just after this enhanced activity we detected a peak at 43\,GHz in the polarized flux density and an increase in the fractional polarization. 

We found the core region to be polarized by about $1\%$ while the jet region showed an average fractional  polarization of $\sim 16\%$. In the jet region EVPAs were roughly orthogonal to the jet position angle, implying a magnetic field parallel to the jet axis. In the core region, the EVPAs varied between $\sim100^\circ$ and $\sim150^\circ$ at 43\,GHz, but they were more variable at lower frequencies, in particular at 15\,GHz, because of opacity effects and variable Faraday rotation. 
Near the VLBI radio core we confirmed the presence of a limb-brightened structure in polarized emission and a transverse EVPA distribution, as previously reported by \citet{Piner2010} and \citet{Giroletti2008}.

Thanks to these multi-frequency and multi-epoch datasets we constrained polarization parameters, and we proposed the following scenario to explain the observed properties: The parallel magnetic field in the jet region could be caused by velocity shear across the jet. These velocity gradients might cause Kelvin-Helmholtz instabilities that would heat the boundary layer, causing the observed limb brightening. This physical mechanism may have been favored by the enhanced activity, that foreshadowed the appearance of the limb brightened structure. 
To explain the intrinsic EVPA variations in the core region, we proposed subcomponents within the VLBI core region, with different polarization properties, which also result in the low degree of polarization. 

The polarimetric information presented in this paper complements the variability analysis discussed in \citet{Lico2012} and in \citet{Blasi2013}. The measurements are from a multi-frequency campaign carried out during 2011, which involves observations throughout the electromagnetic spectrum. For this reason, our results provide a starting point for future broadband analyses that could improve our knowledge about the emission of TeV blazars.

\begin{table*}
\begin{center}
\begin{tiny}
\caption{Summary of the total and polarized intensity parameters plotted in Figs.~\ref{plots_core} and \ref{plots_jet}.}
\label{table_data}   
\begin{tabular}{ccccccccccc}  
\hline\hline                 
Epoch & Frequency & Region & S\tablefootmark{a} & $\sigma_S$\tablefootmark{b} & P\tablefootmark{c} & $\sigma_P$\tablefootmark{d} & m\tablefootmark{e} & $\sigma_m$\tablefootmark{f} & $\chi$\tablefootmark{g} & $\sigma_{\chi}$\tablefootmark{h} \\
year/month/day & (GHz) & & (mJy) & (mJy) & (mJy) & (mJy) & (\%) & (\%) & (deg) & (deg) \\
\hline\hline                 
&&&&&&&&&&\\
2011/01/14 & 15 & Core & 332.2 & 33.2 & 4.9  & 0.5 & 1.5  & 0.2 & 108.3 & 8.7  \\
           &    & Jet  & 60.5  & 6.1  & 8.9  & 1.0 & 14.8 & 2.3 & 58.3  & 7.5  \\
           & 24 & Core & 318.9 & 31.9 & 2.7  & 0.3 & 0.8  & 0.1 & 126.5 & 8.8  \\
           &    & Jet  & 47.4  & 4.8  & 6.4  & 0.9 & 13.6 & 2.3 & 58.5  & 9.0  \\
           & 43 & Core & 291.2 & 29.1 & 2.2  & 0.3 & 0.8  & 0.1 & 117.8 & 10.9 \\
2011/02/25 & 15 & Core & 392.9 & 39.3 & 4.2  & 0.4 & 1.1  & 0.2 & 117.3 & 7.8  \\
           &    & Jet  & 43.4  & 4.4  & 6.5  & 0.8 & 15.0 & 2.5 & 55.3  & 8.1  \\
           & 24 & Core & 419.9 & 42.0 & 4.2  & 0.5 & 1.0  & 0.1 & 143.5 & 8.2  \\
           &    & Jet  & 45.1  & 4.6  & 5.7  & 0.9 & 12.6 & 2.3	 & 55.5 & 8.7 \\
           & 43 & Core & 408.6 & 40.9 & 2.6  & 0.3 & 0.6  & 0.1 & 113.8 & 10.2 \\
2011/03/29 & 15 & Core & 399.6 & 40.0 & 2.6  & 0.3 & 0.7  & 0.1 & 177.3 & 12.4 \\
           &    & Jet  & 55.7  & 5.6  & 7.6  & 0.9 & 13.7 & 2.2 & 56.3  & 12.6 \\
           & 24 & Core & 364.7 & 36.5 & 5.5  & 0.6 & 1.5  & 0.2 & 172.5 & 12.3 \\
           &    & Jet  & 39.6  & 4.0  & 11.5 & 1.3 & 29.1 & 4.4	& 61.5  & 12.5 \\
           & 43 & Core & 346.8 & 34.7 & 12.5 & 1.4 & 3.6  & 0.5 & 142.8 & 13.1 \\
2011/04/25 & 15 & Core & 367.4 & 36.7 & 5.1  & 0.5 & 1.4  & 0.2 & 174.3 & 10.5 \\
           &    & Jet  & 47.3  & 4.8  & 16.5 & 1.8 & 34.8 & 5.1 & 58.3  & 8.9  \\
           & 24 & Core & 328.8 & 32.9 & 6.5  & 0.7 & 2.0  & 0.3 & 157.5 & 10.5 \\
           &    & Jet  & 44.0  & 4.5  & 16.4 & 1.7 & 37.3 & 5.5	& 72.5  & 10.5 \\
           & 43 & Core & 306.7 & 30.7 & 8.1  & 0.8 & 2.6  & 0.4 & 130.8 & 11.8 \\
2011/05/31 & 15 & Core & 344.4 & 34.4 & 1.1  & 0.2 & 0.3  & 0.1 & 130.3 & 13.8 \\
           &    & Jet  & 43.7  & 4.4  & 5.4  & 0.8 & 12.3 & 2.2 & 51.3  &  9.6 \\
           & 24 & Core & 301.4 & 30.1 & 3.0  & 0.3 & 1.0  & 0.1 & 168.5 & 10.1 \\
           &    & Jet  & 35.3  & 3.6  & 3.7  & 0.7 & 10.4 & 2.2	& 66.5  & 11.4 \\
           & 43 & Core & 246.8 & 24.7 & 3.4  & 0.4 & 1.4  & 0.2 & 149.8 & 10.1 \\
2011/06/29 & 15 & Core & 264.6 & 26.5 & 1.8  & 0.2 & 0.7  & 0.1 & 102.2 & 11.4 \\
           &    & Jet  & 55.6  & 5.6  & 8.9  & 1.1 & 16.0 & 2.5 & 61.2  & 11.6 \\
           & 24 & Core & 219.5 & 22.0 & 2.2  & 0.3 & 1.0  & 0.2 & 144.0 & 14.7 \\
           &    & Jet  & 37.3  & 3.9  & 6.0  & 1.0 & 16.2 & 3.2	& 46.0  & 13.0 \\
           & 43 & Core & 176.0 & 17.6 & 2.4  & 0.4 & 1.3  & 0.2 & 129.8 & 11.5 \\
2011/07/28 & 15 & Core & 246.8 & 24.7 & 2.6  & 0.4 & 1.1  & 0.2 & 196.2 & 20.2 \\
           &    & Jet  & 45.9  & 4.7  & 6.4  & 1.1 & 13.9 & 2.7 & 56.2  & 20.2 \\
           & 24 & Core & 215.3 & 21.5 & 4.0  & 0.5 & 1.9  & 0.3 & 165.5 & 20.4 \\
           &    & Jet  & 43.3  & 4.4  & 1.8  & 1.1 & 4.2  & 2.5	& 52.5  & 26.3 \\
           & 43 & Core & 195.9 & 19.6 & 3.4  & 0.5 & 1.7  & 0.3 & 159.8 & 20.0 \\
2011/08/29 & 15 & Core & 299.2 & 29.9 & 4.4  & 0.5 & 1.5  & 0.2 & 91.2  & 13.4 \\
           &    & Jet  & 47.3  & 4.8  & 12.6 & 1.5 & 26.6 & 4.2 & 63.2  & 14.4 \\
           & 24 & Core & 259.5 & 25.9 & 2.6  & 0.4 & 1.0  & 0.2 & 115.5 & 13.4 \\
           &    & Jet  & 38.9  & 4.0  & 7.9  & 1.2 & 20.3 & 3.7 & 49.5  & 14.3 \\
           & 43 & Core & 216.1 & 21.6 & 4.7  & 0.5 & 2.2  & 0.3 & 172.8 & 13.7 \\
2011/09/28 & 15 & Core & 275.4 & 27.5 & 2.5  & 0.3 & 0.9  & 0.1 & 181.2 & 18.1 \\
           &    & Jet  & 38.7  & 3.9  & 3.8  & 0.8 & 9.7  & 2.2 & 57.2  & 18.8 \\
           & 24 & Core & 258.7 & 25.9 & 4.0  & 0.5 & 1.6  & 0.2 & 175.5 & 17.9 \\
           &    & Jet  & 42.2  & 4.3  & 5.3  & 1.0 & 12.7 & 2.8 & 42.5  & 19.2 \\
           & 43 & Core & 227.0 & 22.7 & 3.0  & 0.4 & 1.3  & 0.2 & 149.8 & 18.5 \\
2011/10/29 & 15 & Core & 267.8 & 26.8 & 4.7  & 0.5 & 1.8  & 0.3 & 108.2 & 10.5 \\
           &    & Jet  & 53.1  & 5.4  & 5.5  & 0.9 & 10.3 & 2.0 & 65.2  & 11.1 \\
           & 24 & Core & 259.7 & 26.0 & 1.8  & 0.3 & 0.7  & 0.1 & 103.5 & 13.3 \\
           &    & Jet  & 34.1  & 3.5  & 6.8  & 0.9 & 19.9 & 3.4	& 43.5  & 12.1 \\
           & 43 & Core & 233.9 & 23.4 & 2.5  & 0.4 & 1.1  & 0.2 & 140.8 & 11.5 \\
2011/11/28 & 15 & Core & 303.8 & 30.4 & 2.6  & 0.4 & 0.9  & 0.1 & 106.5 & 17.1 \\
           &    & Jet  & 43.1  & 4.4  & 7.0  & 1.1 & 16.2 & 3.1 & 83.5  & 17.6 \\
           & 24 & Core & 304.1 & 30.4 & 2.1  & 0.3 & 0.7  & 0.1 & 109.3 & 18.5 \\
           &    & Jet  & 40.7  & 4.3  & 2.9  & 1.0 & 7.0  & 2.6	& 115.3 & 19.6 \\
           & 43 & Core & 278.5 & 27.9 & 3.1  & 0.4 & 1.1  & 0.2 & 152.8 & 17.2 \\
2011/12/23 & 15 & Core & 303.8 & 30.4 & 2.9  & 0.4 & 1.0  & 0.2 & 114.5 & 14.1 \\
           &    & Jet  & 46.3  & 4.7  & 3.6  & 0.9 & 7.8  & 2.0 & 64.5  & 13.6 \\
           & 24 & Core & 284.9 & 28.5 & 1.9  & 0.3 & 0.7  & 0.1 & 144.3 & 13.8 \\
           &    & Jet  & 25.8  & 2.7  & 0.4  & 0.6 & 1.6  & 2.3 & - & - \\
           & 43 & Core & 202.9 & 20.3 & 3.3  & 0.4 & 1.6  & 0.3 & 162.8 & 12.5 \\
\hline                                  
\end{tabular}
\end{tiny}
\tablefoot{
\begin{tiny}
\newline
\tablefoottext{a}{Flux density in mJy.}\\
\tablefoottext{b}{Estimated errors for the flux density.}\\
\tablefoottext{c}{Polarized flux density in mJy.}\\
\tablefoottext{d}{Estimated errors for the polarized flux density.}\\
\tablefoottext{e}{Fractional polarization.}\\
\tablefoottext{f}{Estimated errors for the fractional polarization.}\\
\tablefoottext{g}{EVPAs.}\\
\tablefoottext{h}{Estimated errors for the EVPA.}
\end{tiny}
}
\end{center}
\end{table*}

\begin{acknowledgement}
\begin{small}
The \textit{Fermi} LAT Collaboration acknowledges generous ongoing support
from a number of agencies and institutes that have supported both the
development and the operation of the LAT as well as scientific data analysis.
These include the National Aeronautics and Space Administration and the
Department of Energy in the United States, the Commissariat \`a l'Energie Atomique
and the Centre National de la Recherche Scientifique / Institut National de Physique
Nucl\'eaire et de Physique des Particules in France, the Agenzia Spaziale Italiana
and the Istituto Nazionale di Fisica Nucleare in Italy, the Ministry of Education,
Culture, Sports, Science and Technology (MEXT), High Energy Accelerator Research
Organization (KEK) and Japan Aerospace Exploration Agency (JAXA) in Japan, and
the K.~A.~Wallenberg Foundation, the Swedish Research Council and the
Swedish National Space Board in Sweden.
 
Additional support for science analysis during the operations phase is gratefully
acknowledged from the Istituto Nazionale di Astrofisica in Italy and the Centre National d'\'Etudes Spatiales in France. \\
This work is based on observations obtained through the BG207 VLBA project, which makes use of the Swinburne University of Technology software correlator, developed as part of the Australian Major National Research Facilities Programme and operated under licence \citep{Deller2011}. The
National Radio Astronomy Observatory is a facility of the National Science Foundation operated
under cooperative agreement by Associated Universities, Inc. For this paper we made use of the NASA/IPAC Extragalactic Database NED which is operated by the JPL, Californian Institute of Technology, under contract with the National Aeronautics and Space Administration. We acknowledge financial contribution from grant PRIN-INAF-2011. KVS and YYK are partly supported by the Russian Foundation for Basic Research (project 13-02-12103). KVS is also supported by the RFBR grant 14-02-31789. YYK is also supported by the Dynasty Foundation. The research at Boston University was supported in part by NASA through Fermi grants NNX08AV65G, NNX08AV61G, NNX09AT99G, NNX09AU10G, and NNX11AQ03G, and by US National Science Foundation grant AST-0907893. Part of this work was supported by the Marco Polo program of the University of Bologna and the COST Action MP0905 "Black Holes in a Violent Universe". The research at the Instituto de Astrofisica de Andalucia was supported in part by the Spanish Ministry of Economy and Competitiveness grant AYA2010-14844 and by the Regional Government of Andalucía (Spain) grant P09-FQM-4784.
We thank Dr. Astrid Peter for the language editing work which improved the text of the present manuscript and the anonymous referee for the valuable comments and suggestions.

\end{small}
\end{acknowledgement}

\end{document}